\documentclass[preprint,10pt]{elsarticle}

\usepackage{amssymb}

\usepackage{amsmath}

\usepackage{soul,color}

\usepackage[acronyms,nonumberlist,nopostdot,nomain,nogroupskip,acronymlists={hidden}]{glossaries}
\newacronym{rmr}{RMR}{RIC Message Router}
\newacronym{3gpp}{3GPP}{3rd Generation Partnership Project}
\newacronym{4g}{4G}{4th generation}
\newacronym{5g}{5G}{5th generation}
\newacronym{6g}{6G}{6th generation}
\newacronym{5gc}{5GC}{5G Core}
\newacronym{adc}{ADC}{Analog to Digital Converter}
\newacronym{aerpaw}{AERPAW}{Aerial Experimentation and Research Platform for Advanced Wireless}
\newacronym{af}{AF}{Application Function}
\newacronym{ai}{AI}{Artificial Intelligence}
\newacronym{aimd}{AIMD}{Additive Increase Multiplicative Decrease}
\newacronym{am}{AM}{Acknowledged Mode}
\newacronym{amc}{AMC}{Adaptive Modulation and Coding}
\newacronym{amf}{AMF}{Access and Mobility Management Function}
\newacronym{an}{AN}{Access Network}
\newacronym{aops}{AOPS}{Adaptive Order Prediction Scheduling}
\newacronym{aoa}{AoA}{Angle of Arrival}
\newacronym{aod}{AoD}{Angle of Departure}
\newacronym{api}{API}{Application Programming Interface}
\newacronym{apn}{APN}{Access Point Name}
\newacronym{aqm}{AQM}{Active Queue Management}
\newacronym{arc}{ARC}{Aerial RAN CoLab}
\newacronym{arc-ota}{ARC-OTA}{Aerial RAN CoLab Over-The-Air}
\newacronym{ausf}{AUSF}{Authentication Server Function}
\newacronym{avc}{AVC}{Advanced Video Coding}
\newacronym{awgn}{AGWN}{Additive White Gaussian Noise}
\newacronym{balia}{BALIA}{Balanced Link Adaptation Algorithm}
\newacronym{bbu}{BBU}{Base Band Unit}
\newacronym{bdp}{BDP}{Bandwidth-Delay Product}
\newacronym{ber}{BER}{Bit Error Rate}
\newacronym{bf}{BF}{Beamforming}
\newacronym{bler}{BLER}{Block Error Rate}
\newacronym{brr}{BRR}{Bayesian Ridge Regressor}
\newacronym{bsr}{BSR}{Buffer Status Report}
\newacronym{bs}{BS}{Base Station}
\newacronym{bpsk}{BPSK}{Binary Phase-shift keying}
\newacronym{bss}{BSS}{Business Support System}
\newacronym{ca}{CA}{Carrier Aggregation}
\newacronym{caas}{CaaS}{Connectivity-as-a-Service}
\newacronym{cb}{CB}{Code Block}
\newacronym{cc}{CC}{Congestion Control}
\newacronym{ccid}{CCID}{Congestion Control ID}
\newacronym{cco}{CC}{Carrier Component}
\newacronym{cdd}{CDD}{Cyclic Delay Diversity}
\newacronym{cdf}{CDF}{Cumulative Distribution Function}
\newacronym{cdn}{CDN}{Content Distribution Network}
\newacronym{ci}{CI}{Continuous Integration}
\newacronym{cicd}{CI/CD}{Continuous Integration/Continuous Delivery}
\newacronym{cir}{CIR}{Channel Impulse Response}
\newacronym{cla}{CLA}{Contributor License Agreement}
\newacronym{cn}{CN}{Core Network}
\newacronym{codel}{CoDel}{Controlled Delay Management}
\newacronym{comac}{COMAC}{Converged Multi-Access and Core}
\newacronym{cord}{CORD}{Central Office Re-architected as a Datacenter}
\newacronym{cornet}{CORNET}{COgnitive Radio NETwork}
\newacronym{cosmos}{COSMOS}{Cloud Enhanced Open Software Defined Mobile Wireless Testbed for City-Scale Deployment}
\newacronym{cots}{COTS}{Commercial Off-the-Shelf}
\newacronym{cp}{CP}{Control Plane}
\newacronym{cpu}{CPU}{Central Processing Unit}
\newacronym{cqi}{CQI}{Channel Quality Information}
\newacronym{cr}{CR}{Cognitive Radio}
\newacronym{cran}{CRAN}{Cloud \gls{ran}}
\newacronym{crs}{CRS}{Cell Reference Signal}
\newacronym{csi}{CSI}{Channel State Information}
\newacronym{csirs}{CSI-RS}{Channel State Information - Reference Signal}
\newacronym{ct}{CT}{Continuous Testing}
\newacronym{cu}{CU}{Central Unit}
\newacronym{d2tcp}{D$^2$TCP}{Deadline-aware Data center TCP}
\newacronym{d3}{D$^3$}{Deadline-Driven Delivery}
\newacronym{dac}{DAC}{Digital to Analog Converter}
\newacronym{dag}{DAG}{Directed Acyclic Graph}
\newacronym{darpa}{DARPA}{Defense Advanced Research Projects Agency}
\newacronym{das}{DAS}{Distributed Antenna System}
\newacronym{dash}{DASH}{Dynamic Adaptive Streaming over HTTP}
\newacronym{dc}{DC}{Dual Connectivity}
\newacronym{dccp}{DCCP}{Datagram Congestion Control Protocol}
\newacronym{dce}{DCE}{Direct Code Execution}
\newacronym{dci}{DCI}{Downlink Control Information}
\newacronym{dcl}{DCL}{Dear Colleague Letter}
\newacronym{dctcp}{DCTCP}{Data Center TCP}
\newacronym{dl}{DL}{Downlink}
\newacronym{dmr}{DMR}{Deadline Miss Ratio}
\newacronym{dmrs}{DMRS}{DeModulation Reference Signal}
\newacronym{dn}{DN}{Data Network}
\newacronym{dpu}{DPU}{Data Processing Unit}
\newacronym{drlcc}{DRL-CC}{Deep Reinforcement Learning Congestion Control}
\newacronym{drs}{DRS}{Discovery Reference Signal}
\newacronym{dt}{DT}{Digital Twin}
\newacronym{du}{DU}{Distributed Unit}
\newacronym{e2e}{E2E}{end-to-end}
\newacronym{e2sm}{E2SM}{E2 Service Model}
\newacronym{e2ap}{E2AP}{E2 Application Protocol}
\newacronym{ecaas}{ECaaS}{Edge-Cloud-as-a-Service}
\newacronym{ecn}{ECN}{Explicit Congestion Notification}
\newacronym{edf}{EDF}{Earliest Deadline First}
\newacronym{eirp}{EIRP}{Effective Isotropic Radiated Power}
\newacronym{em}{EM}{Electro-Magnetic}
\newacronym{embb}{eMBB}{Enhanced Mobile Broadband}
\newacronym{empower}{EMPOWER}{EMpowering transatlantic PlatfOrms for advanced WirEless Research}
\newacronym{enb}{eNB}{evolved Node B}
\newacronym{endc}{EN-DC}{E-UTRAN-\gls{nr} \gls{dc}}
\newacronym{epc}{EPC}{Evolved Packet Core}
\newacronym{eps}{EPS}{Evolved Packet System}
\newacronym{es}{ES}{Edge Server}
\newacronym{etsi}{ETSI}{European Telecommunications Standards Institute}
\newacronym[firstplural=Estimated Times of Arrival (ETAs)]{eta}{ETA}{Estimated Time of Arrival}
\newacronym{ebpf}{eBPF}{Extended Berkeley Packet Filter}
\newacronym{xdp}{XDP}{eXpress Data Path}
\newacronym{eutran}{E-UTRAN}{Evolved Universal Terrestrial Access Network}
\newacronym{faas}{FaaS}{Function-as-a-Service}
\newacronym{fapi}{FAPI}{Functional Application Platform Interface}
\newacronym{fcc}{FCC}{Federal Communications Commission}
\newacronym{fdd}{FDD}{Frequency Division Duplexing}
\newacronym{fdm}{FDM}{Frequency Division Multiplexing}
\newacronym{fdma}{FDMA}{Frequency Division Multiple Access}
\newacronym{fed4fire}{FED4FIRE+}{Federation 4 Future Internet Research and Experimentation Plus}
\newacronym{fir}{FIR}{Finite Impulse Response}
\newacronym{fit}{FIT}{Future \acrlong{iot}}
\newacronym{fpga}{FPGA}{Field Programmable Gate Array}
\newacronym{fr1}{FR-1}{Frequency Range 1}
\newacronym{fr2}{FR-2}{Frequency Range 2}
\newacronym{fr3}{FR-3}{Frequency Range 3}
\newacronym{fs}{FS}{Fast Switching}
\newacronym{fscc}{FSCC}{Flow Sharing Congestion Control}
\newacronym{ftp}{FTP}{File Transfer Protocol}
\newacronym{fw}{FW}{Flow Window}
\newacronym{ga128}{Ga}{Golay Sequence type A}
\newacronym{ge}{GE}{Gaussian Elimination}
\newacronym{geo}{GEO}{Geosynchronous Earth Orbit}
\newacronym{glfsr}{GLFSR}{Galois Linear Feedback Shift Register}
\newacronym{gnb}{gNB}{Next Generation Node B}
\newacronym{gold}{Gold}{Gold}
\newacronym{gop}{GOP}{Group of Pictures}
\newacronym{gpr}{GPR}{Gaussian Process Regressor}
\newacronym{gpu}{GPU}{Graphics Processing Unit}
\newacronym{gtp}{GTP}{GPRS Tunneling Protocol}
\newacronym{gtpc}{GTP-C}{GPRS Tunnelling Protocol Control Plane}
\newacronym{gtpu}{GTP-U}{GPRS Tunnelling Protocol User Plane}
\newacronym{gtpv2c}{GTPv2-C}{\gls{gtp} v2 - Control}
\newacronym{gw}{GW}{Gateway}
\newacronym{harq}{HARQ}{Hybrid Automatic Repeat reQuest}
\newacronym{hetnet}{HetNet}{Heterogeneous Network}
\newacronym{hh}{HH}{Hard Handover}
\newacronym{hol}{HOL}{Head-of-Line}
\newacronym{hqf}{HQF}{Highest-quality-first}
\newacronym{hss}{HSS}{Home Subscription Server}
\newacronym{http}{HTTP}{HyperText Transfer Protocol}
\newacronym{ia}{IA}{Initial Access}
\newacronym{iab}{IAB}{Integrated Access and Backhaul}
\newacronym{ic}{IC}{Incident Command}
\newacronym{ietf}{IETF}{Internet Engineering Task Force}
\newacronym{ifw}{IFW}{Interference Free Window}
\newacronym{imsi}{IMSI}{International Mobile Subscriber Identity}
\newacronym{imt}{IMT}{International Mobile Telecommunication}
\newacronym{iot}{IoT}{Internet of Things}
\newacronym{ip}{IP}{Internet Protocol}
\newacronym{iq}{IQ}{In-phase and Quadrature}
\newacronym{isac}{ISAC}{Integrated Sensing and Communication}
\newacronym{itu}{ITU}{International Telecommunication Union}
\newacronym{json}{JSON}{JavaScript Object Notation}
\newacronym{kpi}{KPI}{Key Performance Indicator}
\newacronym{kpm}{KPM}{Key Performance Measurement}
\newacronym{kvm}{KVM}{Kernel-based Virtual Machine}
\newacronym{leo}{LEO}{Low Earth Orbit}
\newacronym{los}{LOS}{Line-of-Sight}
\newacronym{ls}{LS}{Loosely Synchronised}
\newacronym{lsm}{LSM}{Link-to-System Mapping}
\newacronym{lpp}{LPP}{LTE Positioning Protocol}
\newacronym{lte}{LTE}{Long Term Evolution}
\newacronym{lxc}{LXC}{Linux Container}
\newacronym{m2m}{M2M}{Machine to Machine}
\newacronym{mac}{MAC}{Medium Access Control}
\newacronym{manet}{MANET}{Mobile Ad Hoc Network}
\newacronym{mano}{MANO}{Management and Orchestration}
\newacronym{mc}{MC}{Multi-Connectivity}
\newacronym{mcc}{MCC}{Mobile Cloud Computing}
\newacronym{mchem}{MCHEM}{Massive Channel Emulator}
\newacronym{mcs}{MCS}{Modulation and Coding Scheme}
\newacronym{mec}{MEC}{Multi-access Edge Computing}
\newacronym{mec2}{MEC}{Mobile Edge Cloud}
\newacronym{mfc}{MFC}{Mobile Fog Computing}
\newacronym{mi}{MI}{Mutual Information}
\newacronym{mib}{MIB}{Master Information Block}
\newacronym{miesm}{MIESM}{Mutual Information Based Effective SINR}
\newacronym{mimo}{MIMO}{Multiple Input, Multiple Output}
\newacronym{mgen}{MGEN}{Multi-Generator}
\newacronym{ml}{ML}{Machine Learning}
\newacronym{mlr}{MLR}{Maximum-local-rate}
\newacronym[plural=\gls{mme}s,firstplural=Mobility Management Entities (MMEs)]{mme}{MME}{Mobility Management Entity}
\newacronym{mmtc}{mMTC}{Massive Machine-Type Communications}
\newacronym{mmwave}{mmWave}{millimeter wave}
\newacronym{mpdccp}{MP-DCCP}{Multipath Datagram Congestion Control Protocol}
\newacronym{mptcp}{MPTCP}{Multipath TCP}
\newacronym{mr}{MR}{Maximum Rate}
\newacronym{mrdc}{MR-DC}{Multi \gls{rat} \gls{dc}}
\newacronym{mse}{MSE}{Mean Square Error}
\newacronym{mss}{MSS}{Maximum Segment Size}
\newacronym{mt}{MT}{Mobile Termination}
\newacronym{mtd}{MTD}{Machine-Type Device}
\newacronym{mtu}{MTU}{Maximum Transmission Unit}
\newacronym{mumimo}{MU-MIMO}{Multi-user \gls{mimo}}
\newacronym{mvno}{MVNO}{Mobile Virtual Network Operator}
\newacronym{nalu}{NALU}{Network Abstraction Layer Unit}
\newacronym{nas}{NAS}{Network Attached Storage}
\newacronym{nbiot}{NB-IoT}{Narrow Band IoT}
\newacronym{nef}{NEF}{Network Exposure Function}
\newacronym{nwdaf}{NWDAF}{Network Data Analytics Function}
\newacronym{nfv}{NFV}{Network Function Virtualization}
\newacronym{nfvi}{NFVI}{Network Function Virtualization Infrastructure}
\newacronym{nic}{NIC}{Network Interface Card}
\newacronym{nlos}{NLOS}{Non-Line-of-Sight}
\newacronym{now}{NOW}{Non Overlapping Window}
\newacronym{nrdz}{NRDZ}{National Radio Dynamic Zone}
\newacronym{nsf}{NSF}{National Science Foundation}
\newacronym{nsm}{NSM}{Network Service Mesh}
\newacronym{nr}{NR}{New Radio}
\newacronym{nrf}{NRF}{Network Repository Function}
\newacronym{nrt}{near-RT}{near-Real-Time}
\newacronym{nsa}{NSA}{Non Standalone}
\newacronym{nse}{NSE}{Network Slicing Engine}
\newacronym{nssf}{NSSF}{Network Slice Selection Function}
\newacronym{ntn}{NTN}{Non-Terrestrial Network}
\newacronym{ntp}{NTP}{Network Time Protocol}
\newacronym{o2i}{O2I}{Outdoor to Indoor}
\newacronym{oai}{OAI}{OpenAirInterface}
\newacronym{oaic}{OAIC}{Open AI Cellular}
\newacronym{oaicn}{OAI-CN}{\gls{oai} \acrlong{cn}}
\newacronym{oairan}{OAI-RAN}{\acrlong{oai} \acrlong{ran}}
\newacronym{oam}{OAM}{Operations, Administration and Maintenance}
\newacronym[plural=\gls{obu}s,firstplural=Onboard Units (OBUs)]{obu}{OBU}{Onboard Unit}
\newacronym{ofdm}{OFDM}{Orthogonal Frequency Division Multiplexing}
\newacronym{olia}{OLIA}{Opportunistic Linked Increase Algorithm}
\newacronym{omec}{OMEC}{Open Mobile Evolved Core}
\newacronym{onap}{ONAP}{Open Network Automation Platform}
\newacronym{onf}{ONF}{Open Networking Foundation}
\newacronym{onos}{ONOS}{Open Networking Operating System}
\newacronym{oom}{OOM}{\gls{onap} Operations Manager}
\newacronym{opnfv}{OPNFV}{Open Platform for \gls{nfv}}
\newacronym{orbit}{ORBIT}{Open-Access Research Testbed for Next-Generation Wireless Networks}
\newacronym{os}{OS}{Operating System}
\newacronym{osa}{OSA}{OAI Software Alliance}
\newacronym{osc}{OSC}{O-RAN Software Community}
\newacronym{oss}{OSS}{Operations Support System}
\newacronym{pa}{PA}{Position-aware}
\newacronym{pase}{PASE}{Prioritization, Arbitration, and Self-adjusting Endpoints}
\newacronym{pawr}{PAWR}{Platforms for Advanced Wireless Research}
\newacronym{pbch}{PBCH}{Physical Broadcast Channel}
\newacronym{pci}{PCI}{Peripheral Component Interconnect}
\newacronym{pcf}{PCF}{Policy Control Function}
\newacronym{pcef}{PCEF}{Policy and Charging Enforcement Function}
\newacronym{pcfich}{PCFICH}{Physical Control Format Indicator Channel}
\newacronym{pcrf}{PCRF}{Policy and Charging Rules Function}
\newacronym{pdcch}{PDCCH}{Physical Downlink Control Channel}
\newacronym{pdcp}{PDCP}{Packet Data Convergence Protocol}
\newacronym{pdsch}{PDSCH}{Physical Downlink Shared Channel}
\newacronym{pdu}{PDU}{Packet Data Unit}
\newacronym{pdp}{PDP}{Power Delay Profile}
\newacronym{pf}{PF}{Proportional Fair}
\newacronym{pgw}{PGW}{Packet Gateway}
\newacronym{ph}{PH}{Power Headroom}
\newacronym{phich}{PHICH}{Physical Hybrid ARQ Indicator Channel}
\newacronym{phy}{PHY}{Physical}
\newacronym{pl}{PL}{Path Loss}
\newacronym{pmch}{PMCH}{Physical Multicast Channel}
\newacronym{pmi}{PMI}{Precoding Matrix Indicators}
\newacronym{powder}{POWDER}{Platform for Open Wireless Data-driven Experimental Research}
\newacronym{ppo}{PPO}{Proximal Policy Optimization}
\newacronym{ppp}{PPP}{Poisson Point Process}
\newacronym{prach}{PRACH}{Physical Random Access Channel}
\newacronym{prb}{PRB}{Physical Resource Block}
\newacronym{psnr}{PSNR}{Peak Signal to Noise Ratio}
\newacronym{pss}{PSS}{Primary Synchronization Signal}
\newacronym{pucch}{PUCCH}{Physical Uplink Control Channel}
\newacronym{pusch}{PUSCH}{Physical Uplink Shared Channel}
\newacronym{qam}{QAM}{Quadrature Amplitude Modulation}
\newacronym{qci}{QCI}{\gls{qos} Class Identifier}
\newacronym{qoe}{QoE}{Quality of Experience}
\newacronym{qos}{QoS}{Quality of Service}
\newacronym{qtgui}{QT-GUI}{QT Graphical User Interface}
\newacronym{qsfp28}{QSFP28}{Quad Small Form-factor Pluggable 28}
\newacronym{quic}{QUIC}{Quick UDP Internet Connections}
\newacronym{rach}{RACH}{Random Access Channel}
\newacronym{ran}{RAN}{Radio Access Network}
\newacronym[firstplural=Radio Access Technologies (RATs)]{rat}{RAT}{Radio Access Technology}
\newacronym{rc}{RC}{RAN Control}
\newacronym{rcn}{RCN}{Research Coordination Network}
\newacronym{rec}{REC}{Radio Edge Cloud}
\newacronym{red}{RED}{Random Early Detection}
\newacronym{renew}{RENEW}{Reconfigurable Eco-system for Next-generation End-to-end Wireless}
\newacronym{rf}{RF}{Radio Frequency}
\newacronym{rfc}{RFC}{Request for Comments}
\newacronym{rfr}{RFR}{Random Forest Regressor}
\newacronym{ric}{RIC}{RAN Intelligent Controller}
\newacronym{near-rt-ric}{near-RT-RIC}{near-Real-Time-\gls{ric}}
\newacronym{rlc}{RLC}{Radio Link Control}
\newacronym{rlf}{RLF}{Radio Link Failure}
\newacronym{rlnc}{RLNC}{Random Linear Network Coding}
\newacronym{rmse}{RMSE}{Root Mean Squared Error}
\newacronym{rnis}{RNIS}{Radio Network Information Service}
\newacronym{rr}{RR}{Round Robin}
\newacronym{rrc}{RRC}{Radio Resource Control}
\newacronym{rrm}{RRM}{Radio Resource Management}
\newacronym{rru}{RRU}{Remote Radio Unit}
\newacronym{rs}{RS}{Remote Server}
\newacronym{rsrp}{RSRP}{Reference Signal Received Power}
\newacronym{rsrq}{RSRQ}{Reference Signal Received Quality}
\newacronym{rss}{RSS}{Received Signal Strength}
\newacronym{rssi}{RSSI}{Received Signal Strength Indicator}
\newacronym{rsu}{RSU}{Road-Side Unit}
\newacronym{rtt}{RTT}{Round Trip Time}
\newacronym{ru}{RU}{Radio Unit}
\newacronym{rw}{RW}{Receive Window}
\newacronym{rx}{RX}{Receiver}
\newacronym{s1ap}{S1AP}{S1 Application Protocol}
\newacronym{sa}{SA}{Standalone}
\newacronym{sack}{SACK}{Selective Acknowledgment}
\newacronym{sap}{SAP}{Service Access Point}
\newacronym{sas}{SAS}{Spectrum Access System}
\newacronym{sc2}{SC2}{Spectrum Collaboration Challenge}
\newacronym{scef}{SCEF}{Service Capability Exposure Function}
\newacronym{sch}{SCH}{Secondary Cell Handover}
\newacronym{scoot}{SCOOT}{Split Cycle Offset Optimization Technique}
\newacronym{sfp+}{SFP+}{Small Form-factor Pluggable Plus}
\newacronym{sctp}{SCTP}{Stream Control Transmission Protocol}
\newacronym{scp}{SCP}{Service Communication Proxy}
\newacronym{sdap}{SDAP}{Service Data Adaptation Protocol}
\newacronym{sd}{SD}{Standard Deviation}
\newacronym{sdk}{SDK}{Software Development Kit}
\newacronym{sdm}{SDM}{Space Division Multiplexing}
\newacronym{sdma}{SDMA}{Spatial Division Multiple Access}
\newacronym{sdn}{SDN}{Software-defined Networking}
\newacronym{sdr}{SDR}{Software-defined Radio}
\newacronym{seba}{SEBA}{SDN-Enabled Broadband Access}
\newacronym{sgsn}{SGSN}{Serving GPRS Support Node}
\newacronym{sgw}{SGW}{Service Gateway}
\newacronym{si}{SI}{Study Item}
\newacronym{sib}{SIB}{Secondary Information Block}
\newacronym{sinr}{SINR}{Signal to Interference plus Noise Ratio}
\newacronym{sip}{SIP}{Session Initiation Protocol}
\newacronym{siso}{SISO}{Single Input, Single Output}
\newacronym{sla}{SLA}{Service Level Agreement}
\newacronym{sm}{SM}{Service Model}
\newacronym{smf}{SMF}{Session Management Function}
\newacronym{smo}{SMO}{Service Management and Orchestration}
\newacronym{sms}{SMS}{Short Message Service}
\newacronym{smsgmsc}{SMS-GMSC}{\gls{sms}-Gateway}
\newacronym{snr}{SNR}{Signal-to-Noise-Ratio}
\newacronym{son}{SON}{Self-Organizing Network}
\newacronym{sptcp}{SPTCP}{Single Path TCP}
\newacronym{srb}{SRB}{Service Radio Bearer}
\newacronym{srn}{SRN}{Standard Radio Node}
\newacronym{srs}{SRS}{Sounding Reference Signal}
\newacronym{ss}{SS}{Synchronization Signal}
\newacronym{ssh}{SSH}{Secure Shell}
\newacronym{sss}{SSS}{Secondary Synchronization Signal}
\newacronym{st}{ST}{Spanning Tree}
\newacronym{svc}{SVC}{Scalable Video Coding}
\newacronym{synce}{SyncE}{Synchronous Ethernet}
\newacronym{tb}{TB}{Transport Block}
\newacronym{tcp}{TCP}{Transmission Control Protocol}
\newacronym{tdd}{TDD}{Time Division Duplexing}
\newacronym{tdoa}{TDoA}{Time Difference of Arrival}
\newacronym{tdm}{TDM}{Time Division Multiplexing}
\newacronym{tdma}{TDMA}{Time Division Multiple Access}
\newacronym{tmb}{TMB}{Tower-Mounted Booster}
\newacronym{tfl}{TfL}{Transport for London}
\newacronym{tfrc}{TFRC}{TCP-Friendly Rate Control}
\newacronym{tft}{TFT}{Traffic Flow Template}
\newacronym{tgen}{TGEN}{Traffic Generator}
\newacronym{tip}{TIP}{Telecom Infra Project}
\newacronym{tm}{TM}{Transparent Mode}
\newacronym{to}{TO}{Telco Operator}
\newacronym{toa}{ToA}{Time of Arrival}
\newacronym{tr}{TR}{Technical Report}
\newacronym{trp}{TRP}{Transmitter Receiver Pair}
\newacronym{ts}{TS}{Technical Specification}
\newacronym{tti}{TTI}{Transmission Time Interval}
\newacronym{ttt}{TTT}{Time-to-Trigger}
\newacronym{tx}{TX}{Transmitter}
\newacronym{uas}{UAS}{Unmanned Aerial System}
\newacronym{uav}{UAV}{Unmanned Aerial Vehicle}
\newacronym{udm}{UDM}{Unified Data Management}
\newacronym{udp}{UDP}{User Datagram Protocol}
\newacronym{udr}{UDR}{Unified Data Repository}
\newacronym{ue}{UE}{User Equipment}
\newacronym{ueb}{UEB}{User Equipment Booster}
\newacronym{uhd}{UHD}{\gls{usrp} Hardware Driver}
\newacronym{ul}{UL}{Uplink}
\newacronym{um}{UM}{Unacknowledged Mode}
\newacronym{uml}{UML}{Unified Modeling Language}
\newacronym{upa}{UPA}{Uniform Planar Array}
\newacronym{upf}{UPF}{User Plane Function}
\newacronym{urllc}{URLLC}{Ultra Reliable and Low Latency Communication}
\newacronym{usa}{U.S.}{United States}
\newacronym{udsf}{UDSF}{Unstructured Data Storage Function}
\newacronym{usim}{USIM}{Universal Subscriber Identity Module}
\newacronym{usrp}{USRP}{Universal Software Radio Peripheral}
\newacronym{utc}{UTC}{Urban Traffic Control}
\newacronym{vim}{VIM}{Virtualization Infrastructure Manager}
\newacronym{vlan}{VLAN}{Virtual Local Area Network}
\newacronym{vm}{VM}{Virtual Machine}
\newacronym{vnf}{VNF}{Virtual Network Function}
\newacronym{volte}{VoLTE}{Voice over \gls{lte}}
\newacronym{voltha}{VOLTHA}{Virtual OLT HArdware Abstraction}
\newacronym{vr}{VR}{Virtual Reality}
\newacronym{vran}{vRAN}{Virtualized \gls{ran}}
\newacronym{vss}{VSS}{Video Streaming Server}
\newacronym{wbf}{WBF}{Wired Bias Function}
\newacronym{wf}{WF}{Wired-first}
\newacronym{wi}{WI}{Wireless InSite}
\newacronym{wlan}{WLAN}{Wireless Local Area Network}
\newacronym{pnf}{PNF}{Physical Network Function}
\newacronym{drl}{DRL}{Deep Reinforcement Learning}
\newacronym{mtc}{MTC}{Machine-type Communications}
\newacronym{v2x}{V2X}{Vehicle-to-everything}
\newacronym{cast}{\textit{CaST}}{Channel emulation generator and Sounder Toolchain}
\newacronym{abr}{ABR}{Adaptive Bitrate Streaming}
\newacronym{dsp}{DSP}{Digital Signal Processing}
\newacronym{ota}{OTA}{Over-the-Air}
\newacronym{bom}{BoM}{Bill of Materials}
\newacronym{frand}{FRAND}{Fair, Reasonable, And Non-Discriminatory}
\newacronym{nvipc}{NVIPC}{NVIDIA Inter-Process Communication}
\newacronym{hdr}{HDR}{High Dynamic Range}
\newacronym{ipc}{IPC}{Inter-Process Communication}
\newacronym{uci}{UCI}{Uplink Control Indication}
\newacronym{cbrs}{CBRS}{Citizen Broadband Radio Service}
\newacronym{ptp}{PTP}{Precision Timing Protocol}
\newacronym{scf}{SCF}{Small Cell Forum}
\newacronym{prs}{PRS}{Positioning Reference Signal}
\newacronym{nrppa}{NRPPa}{New Radio Positioning Protocol a}
\newacronym{lmf}{LMF}{Localization Management Function}
\newacronym{sep}{SEP}{Standard Essential Patent}
\newacronym{scs}{SCS}{Sub-Carrier Spacing}
\newacronym{oru}{O-RU}{O-RAN RU}
\newacronym{ack}{ACK}{Acknowledgement}
\newacronym{nack}{NACK}{Negative Acknowledgement}
\newacronym{cuup}{CU-UP}{CU User Plane}
\newacronym{cucp}{CU-CP}{CU Control Plane}
\newacronym{snssai}{S-NSSAI}{Single Network Slice Selection Assistance Information}
\newacronym{ipr}{IPR}{Intellectual Property Rights}
\newacronym{osi}{OSI}{Open Source Initiative}
\newacronym{simd}{SIMD}{Single Instruction Multiple Data}
\newacronym{nf}{NF}{Network Function}
\newacronym{sba}{SBA}{Service Based Architecture}
\newacronym{sbi}{SBI}{Service Based Interface}
\newacronym{lf}{LF}{Linux Foundation}
\newacronym{lfn}{LFN}{Linux Foundation Networking}
\newacronym{poc}{PoC}{Proof of Concept}
\newacronym{rbg}{RBG}{Resource Block Group}
\newacronym{ris}{RIS}{Reconfigurable Intelligent Surfaces}
\newacronym{nssai}{NSSAI}{Network Slice Selection Assistance Information}
\newacronym{sriov}{SR-IOV}{Single Root I/O Virtualization}
\newacronym{osm}{OSM}{Open Source MANO}
\newacronym{tsn}{TSN}{Time Sensitive Networking}
\newacronym{mwc}{MWC}{Mobile World Congress}

\usepackage{tikz}
\usepackage{pgfplots}
\usepackage{glossaries}
\usepackage{soul}
\usepackage{xspace}
\usepackage{ragged2e}
\usepackage{subcaption}
\usepackage{enumitem}
\usepackage[hyphens]{url}

\pgfplotsset{compat=newest}
\pgfplotsset{plot coordinates/math parser=false}
\newlength\fheight
\newlength\fwidth
\usetikzlibrary{plotmarks,patterns,decorations.pathreplacing,backgrounds,calc,arrows,arrows.meta,spy,matrix,scopes}
\usepgfplotslibrary{patchplots,groupplots}
\usepackage{tikzscale}

\journal{Computer Networks}

\begin{document}

\begin{frontmatter}

\title{Driving Innovation in 6G Wireless Technologies:\\ The OpenAirInterface Approach}

\author[eur,neu]{Florian Kaltenberger} 
\author[neu]{Tommaso Melodia} 
\author[osa]{Irfan Ghauri}
\author[neu]{Michele Polese} 
\author[eur]{Raymond Knopp} 
\author[eur]{Tien Thinh Nguyen} 
\author[neu]{Sakthivel Velumani}
\author[neu]{Davide Villa}
\author[neu]{Leonardo Bonati}
\author[osa]{Robert Schmidt} 
\author[osa]{Sagar Arora}
\author[br]{Mikel Irazabal}
\author[eur,br]{Navid Nikaein}

\affiliation[eur]{organization={EURECOM},
            city={Sophia Antipolis},
            country={France}}
\affiliation[neu]{organization={Northeastern University},
            city={Boston},
            state={MA},
            country={USA}}
\affiliation[osa]{organization={OpenAirInterface Software Alliance},
            city={Sophia Antipolis},
            country={France}}     
\affiliation[br]{organization={BubbleRAN},
            city={Sophia Antipolis},
            country={France}}

\begin{abstract}
The development of 6G wireless technologies is rapidly advancing, with the \gls{3gpp} entering the pre-standardization phase and aiming to deliver the first specifications by 2028. This paper explores the \gls{oai} project, an open-source initiative that plays a crucial role in the evolution of 5G and future 6G networks. OAI provides a comprehensive implementation of \gls{3gpp} and O-RAN compliant networks, including \gls{ran}, \gls{cn}, and software-defined \gls{ue} components. This paper details the history and evolution of \gls{oai}, its licensing model, and the various projects under its umbrella, such as \gls{ran}, the \gls{cn}, and the \gls{oam} projects. It also highlights the development methodology, \gls{cicd} processes, and end-to-end systems powered by \gls{oai}. Furthermore, the paper discusses the potential of \gls{oai} for 6G research, focusing on spectrum, reflective intelligent surfaces, and \gls{ai}/\gls{ml} integration. The open-source approach of \gls{oai} is emphasized as essential for tackling the challenges of 6G, fostering community collaboration, and driving innovation in next-generation wireless technologies.
\end{abstract}


\end{frontmatter}

\glsresetall

\section{Introduction}
\label{sec:intro}

The year 2024 marks an important milestone in the development of \gls{6g} mobile communication systems as the \gls{3gpp}, the institution that has standardized mobile communication systems since the 3rd generation, has entered the pre-standardization phase of \gls{6g}. It has committed to deliver the first version of the specifications in Release~21, which is scheduled to be completed by the end of 2028~\cite{Larsson2024}. 

While the exact definition of \gls{6g} is still in flux, a high-level set of use cases and requirements has already been specified by the \gls{itu}~\cite{ITU-R-M.2160-0}. These include going beyond the IMT-2020 requirements, such as improved throughput, latency, reliability, spectral efficiency, to include \gls{isac}, Ubiquitous Connectivity, and \gls{ai}. Moreover, four overarching aspects act as design principles commonly applicable to all usage scenarios: sustainability, connecting the unconnected, ubiquitous intelligence, security, and resilience~\cite{ITU-R-M.2160-0}.

From a technical perspective, the community has so far agreed on only a few high-level design principles. In particular, the 6G \gls{ran} should be backwards compatible with 5G and able to connect to an evolution of the 5G \gls{cn}, 6G should support standalone operation only with spectrum sharing between 5G and 6G, and 6G should include open interfaces to foster a healthy ecosystem~\cite{Larsson2024}. Moreover, 6G shall be able to leverage new spectrum opportunities, especially in the upper-mid band around $6-15$~GHz~\cite{ngmn}. More elaborate studies about the use cases, architectures, and possible candidate technologies for 6G can be found in whitepapers from the O-RAN next generation research group~\cite{O-RAN6G} or the EU flagship project HEXA-X-II~\cite{Kerboeuf2024}.

Another fact about 6G is that it will rely much more on open-source software than previous generations. 5G networks were the first to embrace virtualized deployments and open interfaces, thanks to the \gls{sba} of the \gls{3gpp} \gls{cn} as well as the open \gls{ran} architecture specified by the O-RAN ALLIANCE~\cite{o-ran-webpage}. Moreover, many of these virtualized network functions are readily available as open-source software~\cite{Bhat2021, BONATI2020107516}, bringing unprecedented flexibility to 5G systems, making them accessible to a much broader community of researchers and developers, and enabling rapid prototyping and experimentation. Open source will play an even more important role in the development of 6G, where it can serve as early implementations that inform and influence formal standardization processes, making it easier for the ecosystem to converge to practical and scalable solutions for future networks.

This paper presents \gls{oai}, an open-source project that today allows users to easily deploy an end-to-end \gls{3gpp} and O-RAN-compatible \gls{5g} network based on \gls{cots} hardware and interoperable with a wide range of \glspl{ue}. 
In our previous work~\cite{kaltenberger2020}, we have shown how \gls{oai} has democratized 5G  research through open-source access. 
In this work, we now present the latest advancements in \gls{oai} and our vision for evolving this platform toward \gls{6g}. 
We emphasize the importance of our open-source approach and highlight that the challenges of \gls{6g} can only be tackled by a community effort based on open-source components. We believe that this paper can serve as a reference for the wireless research and development community interested in deploying, extending, and experimenting with end-to-end mobile networks based on \gls{oai}. 

This paper is structured as follows: in the remainder of this section, we give a historical overview of the evolution of the OAI project as well as the OAI public license. In Section~\ref{sec:oai_projects}, we present the current state-of-the-art and roadmaps of the different projects under the \gls{osa} umbrella, i.e., \gls{ran}, \gls{cn}, and \gls{oam}. In Section~\ref{sec:CICD}, we present the community-driven development methodology, the contribution process, as well as the \gls{cicd} process. These processes make OAI stand out from other projects in the field. In Section~\ref{sec:end2end}, we give examples of end-to-end deployments of OAI that showcase the versatility of the software. In Section~\ref{sec:6g}, we give our vision of 6G and show how OAI is already used today to showcase some of the features that might make it into 6G. Finally, we draw our conclusions in Section~\ref{sec:conclusions}.

\subsection{OpenAirInterface History and Evolution}
\label{sec:history}

The \gls{oai} project was founded in the early 2000s by EURECOM, a research center and graduate school in Sophia-Antipolis in the south of France. The goal of the project was to bridge the gap between industry and academia by giving researchers an open platform that was as close as possible to the real \gls{3gpp} cellular network. The platform was based on the \gls{sdr} concept, where all the signal processing runs on a general-purpose processor, such as an Intel x86, with dedicated hardware only for analog blocks, such as up and down-conversion, as well as the \gls{rf} transmission and reception chains. Since at the time this technology was still in its infancy and no off-the-shelf devices were available, EURECOM researchers developed these radio cards themselves~\cite{Zayen2015}. 

The first versions of \gls{oai} were only focused on the \gls{ran} and were not interoperable with any third-party core network or user equipment. \gls{oai} first gained widespread attention with the release of \gls{4g} \gls{lte} software around 2010 that made use of the widely available \gls{usrp} \glspl{sdr} from Ettus Research~\cite{usrp} and that was interoperable with \gls{lte} smartphones as well as a third-party \gls{4g} \gls{epc}~\cite{Nikaein2014}.

This success story also sparked a lot of interest in both academia and industry, which led EURECOM to create the \gls{osa} in 2014 to support the growing developer and user community and streamline contributions to the code. \gls{osa} coordinates software contributions from its contributor community and ensures code quality through a well-established review process and a sophisticated \gls{cicd} process based on a test and integration lab it maintains. This testing and integration leverages software simulators, third-party equipment, as well as an over-the-air indoor and outdoor testing environment. 

\gls{osa} maintains a transparent software development process open to the entire \gls{oai} community and publishes a feature roadmap for each of its active projects. \gls{osa} is governed by its board of directors, which is composed of representatives of its founding member, EURECOM, and its strategic partners, which represent a diverse and balanced group of key equipment manufacturers, operators, silicon vendors, software distributors, and startups~\cite{osa_board_members}. The strategic members and the \gls{osa} for-profit associate partners commit financial resources for \gls{osa} that allow the organization to maintain a core team of engineers managing code development, testing, and integration. Non-profit and academic institutions can become associate partners for free and nonetheless contribute to the development of the \gls{oai} software and use it for their research needs. 

The \gls{osa} delivered the first version of a \gls{5g} \gls{ran} capable to support  \gls{5g} \gls{nsa} in 2020 and a \gls{5g} \gls{cn} and \gls{5g} \gls{ran} capable to support \gls{5g} \gls{sa} in 2021~\cite{kaltenberger2020}. Since then, \gls{osa} and several contributors from academia and industry have been evolving  \gls{oai} constantly, integrating additional interfaces such as \gls{3gpp} F1 and E1 interfaces, the \gls{scf} \gls{fapi}~\cite{SCF2021FAPI}, and the O-RAN 7.2 fronthaul, E2, and O1 interfaces. 
Today, \gls{oai} can be deployed on a variety of platforms ranging from fully virtualized cloud platforms to bare metal. It is interoperable with a wide range of 3rd party hardware and software such as \glspl{ru} and \glspl{sdr}, inline and look-aside hardware accelerators, O-RAN compatible \glspl{ric} and \gls{smo} frameworks, as well as \glspl{cn}. The \gls{oai} software assets also offer a standard-compliant test \gls{ue}, an element of fundamental importance for \gls{6g} experimentation. We discuss the current capabilities of \gls{oai} in detail in Section~\ref{sec:oai_projects}.

The \gls{osa} also maintains a strong relationship to other organizations, such as the O-RAN ALLIANCE~\cite{OSA-ORAN-MoU}, the Linux Foundation~\cite{OSA-LF-MoU}, as well as the Small Cell Forum~\cite{scf} to ensure the seamless integration of the \gls{oai} software into the broader 5G and 6G open-source ecosystem. 

\subsection{The OAI Public License}
One of the major contributions of the \gls{osa} to the open-source community is the introduction of the \gls{oai} Public License.
Cellular systems are collectively standardized by the industrial players in the \gls{3gpp} forum. Each player in this process of building consensus contributes technology components from its \gls{ipr} portfolio. Once procedures are standardized, \gls{3gpp} acknowledges and recognizes the \gls{ipr} contributions by different parties therein. To facilitate the adoption of the standardized technology, the organizations that own these patents collectively agree to make the \gls{ipr} available to other parties through \gls{frand} terms. The global adoption of 4G and 5G technologies is a remarkable evidence of the success of the above-mentioned process. 

On the other hand, the \gls{osi}~\cite{osi-website} that promotes licenses whose well-known examples include the Apache v2.0, MIT, and the GNU General Public License series, calls for free (meaning royalty-free, i.e., free from patents) redistribution of software.\footnote{The first versions of OAI were published under the GNU General Public License version 2~\cite{gnu-gpl-v2}.}

Cognizant of the divide in \gls{3gpp} standardization practices and limitations of the \gls{osi} approach producing open-source implementations for this industry, the \gls{osa}, in collaboration with its \gls{3gpp}-invested partners, drafted the \gls{oai} public license Version 1.1 in 2017~\cite{oai-public-license}. This license is a modified version of Apache v2.0 License, incorporating an additional clause that allows contributing parties to grant patent licenses to third parties. This clause brings clarity to the use of \gls{oai} software and aligns the software (re)-distribution policy to that of the \gls{3gpp}-standardized essential technology, which operates under \gls{frand} terms for commercial usage. 

The OAI Public License therefore enables the holders of \gls{3gpp} \gls{sep} to contribute \gls{3gpp} procedures to the \gls{oai} source code in confidence under terms called out in the license and in alignment with the practices of this industry. These \gls{sep} can be found in the \gls{etsi} intellectual property rights online database~\cite{ETSI2024}.

\subsection{Related Work}

In this subsection, we discuss other open-source projects that are in direct ``competition'' with \gls{oai}, i.e., that also implement different parts of 5G \gls{ran} or \gls{cn}. For a comprehensive overview of other open-source projects related to building a fully virtualized and orchestrated 5G network, please refer to~\cite{BONATI2020107516}. 

The srsRAN project~\cite{srsran} provides an open-source implementation of both \gls{cu} and \gls{du}, which is interoperable with \glspl{sdr} like the \glspl{usrp}, as well as with \glspl{oru}. For the latter, they provide their own fronthaul library and do not rely on the fronthaul library from the \gls{osc} like \gls{oai}. The feature set of srsRAN \gls{cu} and \gls{du} is comparable to that of \gls{oai}, except that they do not (yet) support \gls{cu} control- and user-plane separation, \gls{srs}, and \gls{fr2}. However, the main difference is their licensing model. They follow a dual licensing model where, on one hand, they provide all of their code under the GNU Affero General Public License v3.0 and, on the other hand, they offer a commercial license through their company, called Software Radio Systems. A more detailed comparison of OAI and srsRAN can be found in~\cite{Amini2024, Alves2024}.

On the side of the \gls{cn}, there are two notable open-source projects: Free5GC is a Linux Foundation project published under Apache License 2.0~\cite{free5gc} and Open5GS is backed by NewPlane Inc.\@ and published under the GNU Affero General Public License v3.0 as well as under a commercial license~\cite{open5gs}. In general, \gls{oai} \gls{cn} provides a similar set of features in comparison with these open-source projects., but since \gls{oai} \gls{cn} is strongly used in research projects, it has some advanced features not present in the other projects. For example we  i) provide a high-performance \gls{upf} (\gls{oai} can achieve nearly 100 gigabits per second throughput); ii) implement a cloud-native core with a collection of microservices following open, flexible and extensible principles; iii) support advanced features including \gls{lmf}, \gls{tsn}, iv) providing support for \gls{ai}/\gls{ml}-based Services, enabling  Network Automation for 5G, and providing exposure, and communication enhancements. 

\section{OpenAirInterface Projects}
\label{sec:oai_projects}

\begin{figure}
  \centering
  \includegraphics[width=\columnwidth]{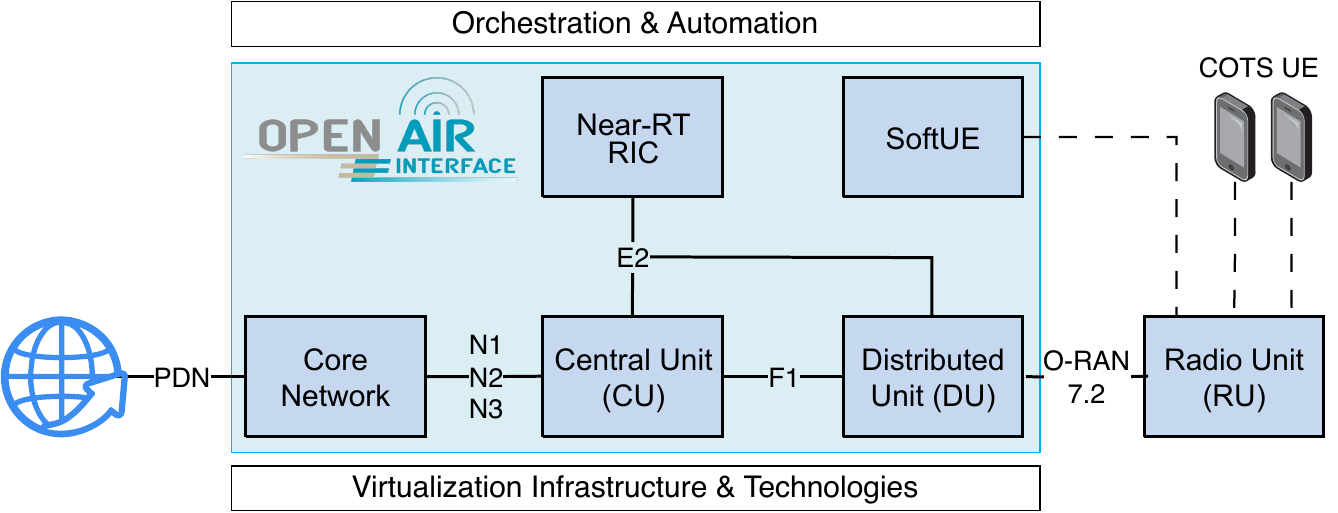}
  \caption{Simplified schematic of a 5G network showing which parts are covered by the \gls{oai} project. All parts of the \gls{oai} software can be run on fully virtualized hardware and can be managed by 3rd party orchestration and automation platforms. }
  \label{fig:cloud_native}
\end{figure}  

Figure \ref{fig:cloud_native} shows a simplified scheme of a 5G network showing the \gls{cn}, the \gls{ran} consisting of \gls{cu}, \gls{du}, and \gls{ru}, the \gls{ric}, as well as the \gls{ue}. 
The components enclosed in the light blue box are the ones covered by the \gls{oai} project, and that we are going to describe in more details in this section.
All parts of the \gls{oai} software can be run on fully virtualized hardware and can be managed by 3rd party orchestration and automation platforms, which we will briefly describe at the end of this section.

The \gls{osa} maintains a roadmap for each of its projects on its webpage~\cite{OSAprojects}. The roadmap is updated approximately every six months and reflects the plan of the development team at the \gls{osa}  as well as commitments from the broader OAI community. The planning horizon is about one year, and the dates on the roadmap are the dates when we expect to merge the feature into the \textit{develop} branch. Most of the features will already be available earlier through the feature branches.

Today, the software supports 4G, 5G \gls{nsa}, as well as 5G \gls{sa} architectures. In the following, however, we focus exclusively on the features of the 5G \gls{sa} implementation.

\subsection{Radio Access Network}
\label{sec:ran}

\begin{figure}
  \centering
  \includegraphics[width=\columnwidth]{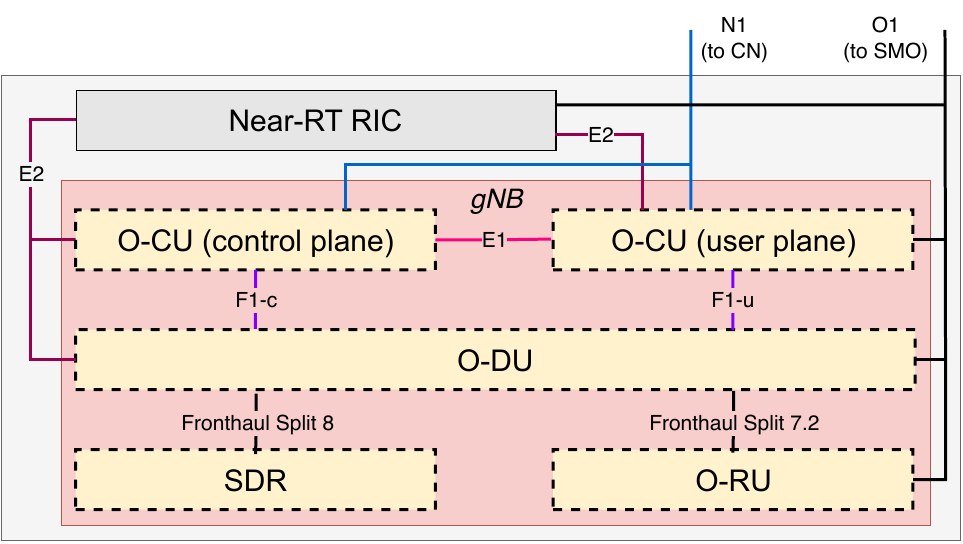}
  \caption{5G RAN architecture including O-RAN interfaces}
  \label{fig:ran}
\end{figure}  

Figure~\ref{fig:ran} shows the basic architecture of a \gls{5g} \gls{ran}  including O-RAN interfaces. The \gls{oai} \gls{ran}  project implements the \gls{cu} and \gls{du} network functions together with all the necessary 3GPP and O-RAN interfaces. \gls{oai} also has an implementation of a  \gls{ric}, which is described in detail in Section~\ref{sec:ric}. The \gls{oai} \gls{ran} can be deployed with the full F1 and E1 splits, i.e., with \gls{cuup}, \gls{cucp}, and \gls{du}, or it can also be configured as a monolithic \gls{gnb}. 

The \gls{oai} \gls{du} implements \gls{rlc}, \gls{mac}, and the \gls{phy} layers of the protocol stack. It can interface with \glspl{sdr} like the \gls{usrp} as well as commercial \glspl{ru}, either using eCPRI split~8 or the O-RAN split~7.2 fronthaul interface, respectively. OAI has been tested successfully with \glspl{oru} from VVDN, Benetel, LiteOn, and Foxconn~\cite{OSA23fronthaul}. 

The \gls{phy} supports static \gls{tdd} as well as \gls{fdd}, subcarrier spacings of 15 kHz, 30 kHz, (\gls{fr1}), and 120 kHz (\gls{fr2}), cell bandwidths of 10, 20, 40, 60, 80, and 100 MHz, as well as 200 MHz (\gls{fr2} only). We support 4-layer \gls{dl} and 2-layer \gls{ul} \gls{mimo} as well as highly efficient 3GPP compliant channel encoder and decoder (turbo, LDPC, polar).

The \gls{phy} can run entirely and in real-time on x86 architecture by leveraging its \gls{simd} vector processing extensions. Layers 2 and above also run on ARM architectures. The \gls{du} can also make use of different kinds of hardware accelerators, both in-line and look-aside. The in-line accelerator makes use of the \gls{fapi} interface between the \gls{phy} and the \gls{mac} layer. Currently, \gls{oai} supports the NVIDIA Aerial \gls{sdk} Layer 1~\cite{villa2024x5g,OSA23ARC}, but others will follow. The look-aside accelerator makes use of the Intel bb-dev interface and currently supports AMD Xilinx T1 and T2 cards~\cite{OSA23lookaside,OSA24lookaside}.

\gls{oai} implements both monolithic \gls{cu} or disaggregated \glspl{cu} (\gls{cuup} and \gls{cucp}). The \gls{cuup} contains mainly the \gls{sdap} and \gls{pdcp} layers. The \gls{cucp} features the \gls{rrc} layer, for radio resource management and \gls{ue} lifecycle management. The communication between the two happens over the E1 interface. Both  \gls{cuup} and  \gls{cucp} can handle multiple \glspl{du}. In addition, F1 handover between multiple \glspl{du} is supported. The handover can either be triggered manually or based on neighbor cell measurements and events. 

Both \gls{oai} \gls{du} and \gls{cu} implement the O-RAN E2 interface to the \gls{nrt} \gls{ric}. The E2 agent supports service models \gls{kpm} v2.03/v3.0 and \gls{rc} v1.03. The E2 interface has been demonstrated successfully with the \gls{osc} \gls{ric}~\cite{Schmidt2023,OSA23OSC-RIC}, as well as with OAI's own implementation of the near \gls{nrt} \gls{ric} called flexRIC~\cite{schmidt21flexRIC,OSA23FlexRIC}.

\gls{oai} \gls{du} and \gls{cu} also implement the O-RAN O1 interface to Operations and Management software such as \gls{onap}~\cite{Kasparik2023, OSA23O1}.

The \gls{oai} \gls{du} and \gls{cu} are compatible with many different open-source core networks such as the Open5GS~\cite{open5gs_website}, free5GC~\cite{free5gc_website}, and of course the OAI 5G Core (see Section~\ref{sec:oai_5gc}) as well as many commercial core network solutions.

For detailed information on the OAI RAN project, we refer the reader to the \textit{readme} file on the OAI git repository~\cite{oai_git}.

\subsubsection{Roadmap}

The following features are part of the roadmap for the RAN project. This implies that early-stage implementation efforts are already underway and will be refined and finalized in the near future.

\paragraph{\gls{ntn}} 
\gls{ntn} refers to the use of space-based infrastructure, such as satellites or high-altitude platforms, to provide 5G wireless communication services. \gls{3gpp} standardized support for \gls{ntn} in Rel.~17. The support for \gls{ntn} in OAI was developed jointly by Fraunhofer IIS, Allbesmart, and University of Luxembourg. It supports both \gls{geo} and \gls{leo} transparent satellites, i.e., the \gls{bs}'s signal is sent from the ground, received and amplified by the satellite before it is sent back to the \gls{ue} on the Earth's surface.  A first demonstration of the developments has been done in a series of papers~\cite{Kumar2022,Kumar2022a,Kumar2022b}. The features are currently being integrated into the main branch of OAI. 

\paragraph{Frequency Range 2 (FR-2)} 
\gls{fr2} refers to deploying the stack with a carrier frequency in the spectrum from $24.25$~GHz to $71$~GHz, also referred to as \gls{mmwave}. When operating in this frequency range, the use of large antenna arrays and beamforming is absolutely necessary to achieve the required link budgets. 3GPP has specified several beamforming procedures for initial beam acquisition, beam tracking, and mobility, as well as beam recovery. Apart from implementing these procedures as well as the other adaptations for \gls{fr2} in \gls{oai}, the most difficult part is getting and setting up the necessary hardware.  A first demonstration of \gls{oai} in \gls{mmwave} was done at \gls{mwc} 2024 with an \gls{usrp} X410 and a \gls{rf} frontend from Interdigital. While this frontend is no longer commercially available, similar performance can be achieved with the  \gls{mmwave} \gls{rf} frontend from TMYTEK~\cite{TMYTEK2024}. At \gls{mwc} 2025 we also demonstrated the integration of \gls{oai} with a \gls{fr2} \gls{oru} from LiteOn.  

\paragraph{Network Slicing and Dynamic \gls{qos} Support} 
As described in Section~\ref{sec:oai_5gc}, the \gls{oai} \gls{cn} already supports Network Slicing and \gls{qos}. At the level of the RAN, these features mostly affect the scheduler. While \gls{qos} ensures optimal performance for various types of data traffic or services by means of prioritization, network slicing tries to create orthogonal resources for different types of data traffic or services, and each slice employs its own scheduler. Orthogonality can be achieved, for example, through allocating different sets of resource blocks or bandwidth parts to different slices. 

\gls{oai} already supports the configuration of multiple slices using the \gls{snssai} and \glspl{ue} can select the desired slice in the \gls{pdu} session establishment. A preliminary version of the scheduler that allows for a static assignment of \glspl{prb} to slices has been shown in~\cite{Cheng2024a} and is currently being integrated into \gls{oai}. A scheduler that supports \gls{qos} has been developed in~\cite{5G-OPERA_D52} and is also being integrated into \gls{oai} soon. 

\paragraph{Geolocalization} 
5G NR supports a wide range of positioning techniques, both network-based and user-based. One of the most common techniques is \gls{ul}-\gls{tdoa}, which is based on synchronized measurements of the \gls{toa} at different \glspl{trp} based on the \gls{srs}  \gls{ul} signal.  These \gls{ul} \gls{toa} measurements are then sent via the \gls{nrppa} protocol to the \gls{lmf}, which performs the estimation of the \gls{ue} position (see Section~\ref{sec:oai_5gc}). A first demonstration of this can be found here~\cite{malik2024concept}.  

The \gls{gnb} also supports the transmission of \gls{prs} signals on the \gls{dl} for \gls{ue}-based positioning (see Section~\ref{sec:soft-ue}).

\paragraph{Sidelink}
5G NR Sidelink refers to a direct communication interface in the 5G New Radio (NR) standard that enables devices to communicate with each other directly through a set of new physical channels. Two independent implementations of these features in OAI were developed in~\cite{Elkadi2023} and~\cite{Mehlhose2024Sidelink}, and OAI is currently working on merging these features and integrating them.

\paragraph{Massive MIMO}
Today, OpenAirInterface supports \gls{mimo} with up to 4 layers in \gls{dl} and 2 layers in \gls{ul}. As mentioned above, OAI is also working on the integration of beamforming procedures. However, OAI does not yet support multi-user MIMO, i.e., spatially separating multiple users on the same time-frequency resources by means of beamforming or precoding. This is usually achieved by means of massive MIMO O-RUs with 32 or more TX and RX chains. These RUs can be operated either on a set of pre-defined beams or by providing weights for the beams that the DU determines based on UL channel measurement (channel reciprocity). While the first one is easier to implement, the latter provides better performance. OAI will develop work to support Massive MIMO in the OAI O-DU solution, including computation of beam weights based on channel reciprocity, \gls{mac} scheduling, as well as support for in-line acceleration through the FAPI interface.

\subsection{FlexRIC}
\label{sec:ric}

FlexRIC~\cite{schmidt21flexRIC} is an open-source implementation of a nearRT-RIC, E2 Agent, and xApp \gls{sdk}. 
It supports \gls{e2ap} v1.0/2.0/3.0 and O-RAN \glspl{sm} i.e., \gls{kpm} v2.01/v2.03/v3.0 and \gls{rc} v1.3, as well as \textit{à la carte} defined \glspl{sm} i.e., MAC, RLC, PDCP and GTP.
Additionally, xApps can be developed using C/C++ and Python3, easing developer experience, and the provided \gls{sdk} already integrates an SQLite database to facilitate further data analysis.
FlexRIC can run on ARM and x86 architectures, as most of its code is strict C11. 
Moreover, its one-way delay latencies from the RAN to the xApp are below 1 ms, which makes it suitable for real-time scenarios, e.g., traffic flow manipulation.
In fact, the importance of traffic control within the RAN has already been demonstrated using \textit{à la carte} TC SM~\cite{irazabal2023tcRAN} on top of OAI, showing FlexRIC project capabilities to handle messages below one millisecond, demonstrating its suitability for real-time scenarios.
Contrary to other projects that sacrifice standard compatibility aiming to achieve low-latency~\cite{foukas23janus, Ko2024edgeric}, FlexRIC is 100\% O-RAN compliant, and its interoperability has been shown with OAI RAN, OSC RIC, Keysight RICTest~\cite{Vladic2023, Schmidt2023}, as well as srsRAN~\cite{srsRAN2024}. 
Lastly, due to FlexRIC's O-RAN standard implementation completeness and thanks to OAI's permissive liberal Public License, its \glspl{sm} have been ported to other nearRT-RIC platforms~\cite{Feraudo2024} and integrated into other testing platforms~\cite{Yonan2024}.

\subsection{User Equipment}
\label{sec:soft-ue}

For a long time, the \gls{oai} \gls{ue} implementation was primarily a tool for testing the \gls{oai} \gls{ran} and core network, so its functionality has been limited to the features specifically required by the \gls{ran}. Moreover, little effort was made to ensure the \gls{oai} \gls{ue} was as stable as the \gls{ran}. However, this has changed in recent years, and the \gls{ue} has undergone significant improvements. Indeed, the software-defined \gls{ue} plays a key role in testing algorithms and protocol stack improvements that span both the \gls{ran} and the \gls{ue} side. Further, it allows for the deployment of software-defined end-to-end cellular networks in a variety of testbeds and experimental environments, including Colosseum, the world's largest wireless network emulator with \glspl{sdr} in the loop. A complete list of features supported by \gls{oai} \gls{ue} can be found here~\cite{oaifeatures}.

One of the main recent achievements has been to make the \gls{oai} \gls{ue} compatible with third-party \glspl{gnb}~\cite{Mongazon2024}. In this demonstration, we used a Nokia \gls{gnb} and an \gls{oai} soft-\gls{ue} with a \gls{usrp} B210 as a radio frontend. One of the main challenges was managing the latency in the \gls{oai} \gls{ue} inherent to the \gls{usrp}. The latency between \gls{dl} and \gls{ul} processing restricts how quickly the \gls{oai} \gls{ue} can send an \gls{ack}/\gls{nack} after a \gls{pdsch} reception (parameter $k_1$~\cite{3gpp38.213}), or how fast the \gls{ue} can transmit on the \gls{pusch} after receiving a \gls{ul} \gls{dci} on the \gls{pdcch} (parameter $k_2$~\cite{3gpp38.214}). Most commercial \glspl{gnb} are configured with short default values for these parameters, typically just 2 slots. However, setting these values to 4 slots allowed the UE to successfully attach to the gNB and initiate a PDU session. 

\subsubsection{Roadmap}

The roadmap for the \gls{oai} \gls{ue} has two main objectives. First, we aim to improve both latency and throughput. Second, we are introducing new capabilities to bring the \gls{ue}'s feature set on par with that of the \gls{gnb}. 

\paragraph{Reducing Latency}

To be able to support $k_1$ and $k_2$ values (see Section~\ref{sec:soft-ue}) of less than 4 slots, the latency in the \gls{ue} has to be reduced. While the latency introduced by the \gls{usrp} is fixed, we can make the processing more efficient by restructuring the \gls{phy} layer to shorten the processing time of physical signals. The main idea is to process received signals on a symbol basis rather than on a slot basis at the \gls{phy} layer. This restructuring would allow the \gls{phy} layer to process \gls{dci} much faster, without waiting for the full slot duration. The same principle can be applied to other \gls{phy} signals. Combined with the recent optimization in \gls{pusch} signal generation, experiments have shown that after restructuring, the minimum $k_1$ and $k_2$ values for normal \gls{ue} operation were reduced by 1 slot, bringing them to 2 slots for $30$~kHz \gls{scs}.

\paragraph{\gls{ntn}}
As mentioned in Section~\ref{sec:ran}, \gls{3gpp} Rel 17 \gls{ntn} are currently being integrated into the \gls{oai} \gls{ran}. Due to the lack of commercial \glspl{ue} supporting \gls{ntn}, an implementation of the corresponding features in the \gls{oai} \gls{ue} is also available. This functionality was demonstrated successfully with a real \gls{geo} satellite~\cite{Kumar2022b,Kumar2022a}, as well as with a \gls{leo} one~\cite{Kumar2023}. 

\paragraph{Geolocalization}
3GPP Release 16 introduced the \gls{prs}, which allows users to determine their location autonomously by using a \gls{dl}-\gls{tdoa} method. The HOP-5G project~\cite{PeralRosado2024,PeralRosado2024a} sponsored the development of the \gls{prs} in \gls{oai} for \gls{dl}-\gls{tdoa} positioning for FR1 and FR2, demonstrating the feasibility of these techniques. Based on this work, \cite{Palama2023} built a testbed with three \gls{oai} \glspl{gnb} and one \gls{oai} \gls{ue} leveraging the \gls{prs} transmissions for \gls{dl}-\gls{tdoa} positioning. More recently, a novel framework presented in~\cite{Mundlamuri2024} estimated the \gls{rtt} between \gls{ue} and \gls{gnb} with existing 5G NR signals.

\paragraph{\gls{ran} Slicing}
Another use case for the \gls{oai} \gls{ue} is \gls{ran} slicing, where the \gls{ue} can handle multiple \gls{pdu} sessions across different slices. Users can interact with the \gls{ue} through the AT command interface to define \gls{pdu} contexts with specific \gls{snssai} values, and request the network to establish \gls{pdu} sessions, similar to commercial serial port-based 5G modems. This functionality

is particularly useful for testing \gls{ran} slicing in controlled environments like the \textit{Colosseum} testbed. Moreover, many 5G \gls{cots} \glspl{ue} do not yet support multiple slices.

\subsection{Core Network}
\label{sec:oai_5gc}
        
\begin{figure}[ht]
\centering
\includegraphics[width=0.99\linewidth]{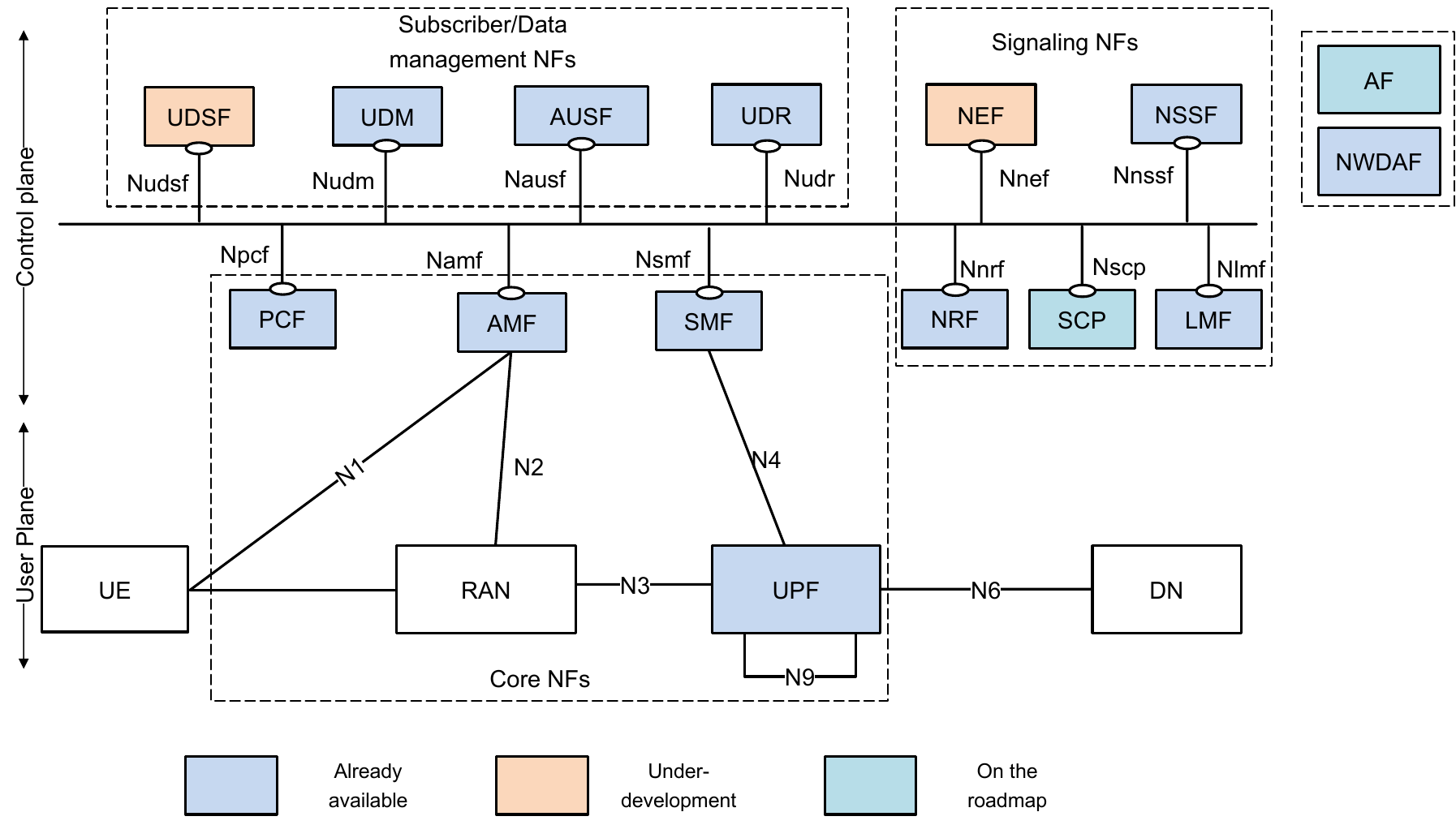}
\caption{OAI 5GC Compoments}
\label{fig:oai_cn_components}
\end{figure}

Figure~\ref{fig:oai_cn_components} depicts the 5G system architecture with its main \glspl{nf}. The 5G system architecture relies on the \gls{sba} to make the functional architecture more suitable for cloud deployment and to provide a flexible framework to meet the needs of a variety of applications and service types in 5G. In the \gls{sba}, the system functionality is achieved by a set of \glspl{nf} providing services to other authorized \glspl{nf} to access their services through a \gls{sbi}. In other words, the \glspl{nf} communicate with each other via \gls{sbi} using cloud-friendly protocols such as HTTP/2 for transport and \gls{json} as the serialization protocol. The centerpiece of the \gls{sba} is the \gls{nrf}, which provides basic \gls{nf} services including authorization, registration/deregistration, and discovery.

The OAI \gls{5gc} complies with 3GPP specifications and offers valuable propositions beyond the standards. Figure \ref{fig:oai_cn_components} shows the current status of OAI \gls{5gc} components from the latest release, rel 2.1.0. In this release, OAI \gls{5gc} supports 11 \glspl{nf}, including \gls{amf}, \gls{smf}, \gls{upf}, \gls{ausf}, \gls{udm}, \gls{udr}, \gls{nrf}, \gls{nssf}, \gls{pcf}, \gls{nwdaf}, and \gls{lmf}. Two \glspl{nf} are currently under development, including \gls{nef} and \gls{udsf}, and additional \glspl{nf}, including \gls{scp} and \gls{af}, will be integrated into the OAI 5GC ecosystem in the upcoming releases.

OAI 5GC supports fundamental 3GPP procedures, including connection, registration, and mobility management procedures (e.g., registration management and service request procedures), as well as session management and N4-related procedures. These capabilities enable the OAI 5GC to support multiple \glspl{ue} attaching to the network and establishing multiple \gls{pdu} sessions concurrently. Additionally, OAI 5GC provides basic functionality to support the \gls{sba}, including NF registration and discovery.

The OAI 5GC also supports several advanced 3GPP procedures and features, including: (i) N2 handover procedures; (ii) event exposure services, which enable third-party applications to collect \gls{ue} statistics and dynamically configure the OAI 5GC based on the collected information via customized APIs; (iii) network slicing-related procedures (OAI 5GC supports network slicing per UE and per PDU session); (iv) support for multiple \glspl{upf} within the same data path, along with the \gls{ul} Classifier feature. These capabilities enable selective traffic routing to a \gls{dn} located near the \gls{an} serving the \gls{ue} and support service and session continuity mode~3.

Flexibility is a key driver of 5G adoption. OAI 5GC provides different flavors of 5GC for both the control and user planes, targeting different 5G applications and use cases. For the user plane, OAI 5GC provides different \gls{upf} implementations. Among them, we highlight:
(i) the simple-switch \gls{upf}, which is easy to deploy and manage, offering sufficient performance for most use cases and therefore well-suited for tasks such as functional testing and validation;
(ii) the \gls{ebpf}/\gls{xdp}-based UPF, a soft-UPF designed to provide a high-performance data path to meet the requirements of various 5G and beyond scenarios. Deployment flexibility is also a key consideration. OAI 5GC can be deployed in a variety of environments, ranging from traditional servers or \glspl{vm} to cloud-native platforms using tools such as Docker Compose or Helm charts.

More advanced features include:
\begin{itemize}
\item Network data analytics and data collection. The NWDAF collects data from various \glspl{nf} and performs analytics to provide insights into the performance and health of the network. Combined with event exposure services, \gls{nwdaf} can serve as a solid foundation for \gls{ai}/\gls{ml}-based decisions to enable automated deployment, management, and optimization of mobile networks.

\item Support for location services. The \gls{lmf} relies on measurements and assistance information from the RAN (via the \gls{nrppa} protocol) and the UE (via the \gls{lpp} protocol) to compute the position of the UE. The first version of the \gls{lmf} includes algorithms to support \gls{ul}-\gls{tdoa} localization based on \gls{toa} measurements from multiple \glspl{gnb}. The \gls{amf} has also been updated to support location services management and to facilitate the transport of location-related messages exchanged between the RAN/UE and the \gls{lmf}.

\item Quality of Service. \gls{qos} refers to the ability to provide differentiated packet forwarding treatment for user data traffic, ensuring specific performance levels such as packet error rate, throughput, delay, and latency. In 5G, the QoS model is based on QoS Flows, which represent the finest granularity of QoS differentiation within a PDU session. To support this architecture, OAI 5GC implements a set of Control Plane features across the \gls{smf}, \gls{pcf}, and \gls{udm}. On the User Plane side, OAI \gls{upf}, leveraging \gls{ebpf}, plays a pivotal role in enforcing QoS and policy rules, offering a promising solution for QoS support in 5G.
\end{itemize}

\subsubsection{Roadmap}

For the upcoming releases, the OAI 5G \gls{cn} will continue to advance its cloud-native and automated deployment model by supporting stateless implementations and introducing additional customized APIs. The \gls{scp} will also be integrated into the OAI 5GC ecosystem to enable various modes of inter-NF communication, thereby supporting diverse deployment and operational scenarios. Finally, to lay the groundwork for Beyond-5G and 6G networks, OAI 5G \gls{cn} will focus on key 5GC features, including QoS, network slicing, support for time-sensitive networking, and enhancements in security and reliability across the entire OAI 5GC ecosystem.

\subsection{Management and Orchestration}
\label{sec:mano}

While \gls{oai} provides essential 5G network functionalities, building a 5G network at scale also requires tools for managing cloud-native environments, as well as orchestration and automation of virtual network functions.

\gls{osm}~\cite{etsi_osm}, a network service orchestrator proposed by \gls{etsi}, is a reference design based on \gls{etsi} \gls{nfv} standard and proposes multiple packages to deploy \gls{oai} 5G \gls{cn} and \gls{ran} functions. 

\gls{onap}~\cite{onap} is an end-to-end network service orchestrator that provides capabilities to manage physical, virtual machine-based, and cloud-native network functions. Certain components of \gls{onap} are also used to implement the O-RAN–proposed \gls{smo}. While \gls{onap} focuses on orchestrating and managing network services, Sylva~\cite{sylva} focuses on orchestrating telco-optimized Kubernetes clusters following cloud-native principles. When building a Kubernetes cluster, the specific requirements of 5G network functions, such as the \gls{upf} or \gls{ran}, are taken into account, including support for technologies like \gls{sriov}, real-time kernels, and others. The Nephio project~\cite{nephio} enables not only the orchestration of Kubernetes clusters but also the orchestration of cloud-native network functions. It leverages GitOps principles along with Kubernetes declarative and reconciliation-based approaches to manage both Kubernetes clusters and cloud-native network functions. Together, all three projects enable the orchestration of end-to-end network services and the management of the underlying cloud infrastructure hosting the network functions.

\section{Development Philosophy and Methodology}
\label{sec:CICD}

\subsection{Importance of Open-Source Communities and their Collaboration}

\gls{osa}, as the foremost open-source community in cellular wireless, understands its leadership role in promoting collaboration. Open source developments in the core technologies (RAN, \gls{cn}, OAM software) of the cellular wireless industry suffer from a variety of challenges namely: (i) the need to maintain an expensive lab for testing and integration of software; (ii) the dearth of highly qualified engineers educated to MSc or PhD level; (iii) cut-throat competition between industry players making direct collaboration unsuitable. Despite these challenges, the value created by open-source reference implementations of a cellular wireless stack is recognized by all stakeholders in the academic community as well as industry for uses such as research and experimentation tools for future studies of the forthcoming standards. 
Under the circumstances, \gls{osa} has played a front-runner’s role in bringing communities like \gls{etsi}, \gls{lfn}, \gls{osc}, and others together. The \gls{osa} leadership and engineers continuously engage with these bodies in thought leadership forums as well as in showcasing \gls{poc} involving software assets from several communities. 
As 6G experimentation advances, \gls{osa} will play a pivotal role in enabling the sandboxes and experimental testbeds on the path to next-generation standardization.

\subsection{Development Process}
The development process at the \gls{oai} project is designed to facilitate high-quality software production, testing, and deployment for 4G and 5G networks. 
The \gls{oai} project is supported by a large international community, contributing to the development and testing of the software. 
The project emphasizes open collaboration, with resources and code available on GitLab~\cite{oai_git}.

Each of the project groups described in Section~\ref{sec:oai_projects} follows a defined roadmap, with specific objectives and milestones that are published on the OAI website~\cite{OSAprojects}. Moreover, each group holds regular open meetings for all the developers to synchronize their developments.  This structured approach helps in tracking progress and maintaining alignment with overall project goals.

\subsection{Contribution Guidelines}

Before a developer can start contributing, they have to sign the OAI \gls{cla}, which basically states that the contributor adheres to the terms of the OAI public license and that they grant a copyright license of the contribution to the Alliance and to recipients of software distributed by the Alliance. In case the author's work is protected by a patent, the author is prepared to grant a license of these patents on \gls{frand} terms to commercial users of the software.  

The developer then creates the contribution, following the basic coding rules and the workflow described in the contribution guidelines of the OAI git repository~\cite{oai_git}.
Once the developer has completed the contribution, they create a merge request on the OAI GitLab page~\cite{oai_git}. This will trigger both human review and automated \gls{ci} tests to evaluate whether a code contribution is ready to be merged.

\subsection{Continuous Integration, Testing, and Deployment}
\label{sec:cicd}

The OAI project leverages a Jenkins-based \gls{cicd} framework, which runs a set of pipelines every time a user creates a merge request on Eurecom's Gitlab server. There are two sets of pipelines:  one set for building images and another set for running tests. The image build pipelines compile the code and also perform static code analysis to catch issues early. The test pipelines run a number of tests, mostly on the various testbenches available at the Eurecom Open5Glab (see Section \ref{sec:open5glab}) but also on external testbenches such as Colosseum (see Section \ref{sec:colosseum}).  

Most of the tests involve \gls{ota} testing with either \gls{cots} \glspl{ue} such as Quectel or with \gls{ue} testers such as the Amarisoft \gls{ue} tester. These tests are complemented with simulation tests, which can either be unitary tests of a specific feature or end-to-end tests using \gls{oai} \gls{rf} simulator. We test 4G, 5G NSA, and 5G SA configurations using either the OAI 4G or 5G \gls{cn}, as well as third-party \glspl{cn}. The tests typically include attach and detach procedures, along with basic connectivity checks using tools such as ping and iPerf. More details can be found in~\cite{OSAtestbenches}.

As an example for an external testbench, Colosseum has been integrated with the OAI \gls{cicd} pipelines to trigger automated tests on the channel emulator, and report results back to the \gls{cicd}. Tests can set different test parameters, such as the OAI branch to test, as well as the RF scenario to emulate during the test. Parameters are passed to an Ansible instance within Colosseum that builds container images for the specified OAI gNB and UE, and benchmarks the performance of the softwarized protocol stacks, leveraging Colosseum automated batch jobs and USRP X310 SDRs as RF transceivers. Once tests are complete, results are processed by Colosseum and summarized into a test report page, which also compares the most recent test with the history of results built over time, and can determine the final outcome of the test (e.g., test passed or failed) based on some criterion set by the developers. Finally, test report, results, and logs are stored in the EURECOM \gls{cicd}, while the test history is maintained on the OpenRAN Gym website~\cite{openrangymwebsite}.

About every week, all the merge requests that have been accepted are combined into an integration branch. The integration branch is tested again with the \gls{ci} pipeline, and if successful, it is merged into \textit{develop} branch and tagged with the year and week number (\textit{year.week}). At the same time, the Docker images according to that tag are pushed to Docker Hub~\cite{osa-dockerhub}. 

\subsection{Performance}

The performance achievable with \gls{oai} depends on a variety of factors. These include the configuration of 5G NR parameters, such as bandwidth and \gls{tdd} settings, as well as the type of radio hardware used, whether \glspl{sdr} or \glspl{oru}. The choice of \gls{ue}, whether \gls{cots} devices or those based on \gls{sdr} and \gls{oai}, also plays a significant role. Additionally, performance is influenced by channel conditions and the number of users connected to the system.

\begin{figure}
\centering
    \setlength\fwidth{0.95\linewidth}
    \setlength\fheight{.3\linewidth}
    \begin{tikzpicture}
\pgfplotsset{every tick label/.append style={font=\scriptsize}}
\definecolor{crimson2143940}{RGB}{214,39,40}
\definecolor{darkgray176}{RGB}{176,176,176}
\definecolor{darkorange25512714}{RGB}{255,127,14}
\definecolor{forestgreen4416044}{RGB}{44,160,44}
\definecolor{lightgray204}{RGB}{204,204,204}
\definecolor{steelblue31119180}{RGB}{31,119,180}

\begin{axis}[
    width=0.8\textwidth,
    height=4.5cm,
    axis lines=box,
    ybar,
    bar width=0pt, 
    xmin=-0.5, xmax=5.5,
    xtick={0,1,2,3,4,5},
    xticklabels={1-fxn-arc,2-fxn-arc,1-vvdn-oai,1-bntl-oai,1-sim-arc,1-sdr-oai},
    xticklabel style={rotate=45, anchor=east},
    xlabel style={font=\scriptsize},
    ylabel={DL Throughput [Gbps]},
    ymin=0, ymax=1.75,
    ytick={0,0.25,0.50,0.75,1.00,1.25,1.50,1.75},
    ylabel style={color=steelblue31119180,font=\scriptsize},
    yticklabel style={color=steelblue31119180},
    tick align=inside,
    tick pos=left,
    ymajorgrids,
    xmajorgrids,
    legend style={
       font=\scriptsize,
       fill opacity=1,
       draw opacity=1,
       text opacity=1,
       draw=black,
       at={(0.15,0.78)},
       anchor=south,
       legend columns=2
    },
    legend cell align={left},
    xlabel={Number of UEs and Network Stack},
    x label style={at={(axis description cs:0.5,-0.4)},anchor=north}
]

\addlegendimage{ybar,ybar legend,draw=black,fill=steelblue31119180,postaction={pattern=north east lines,pattern color=black}}
\addlegendentry{DL}
\addlegendimage{ybar,ybar legend,draw=black,fill=darkorange25512714,postaction={pattern=north west lines,pattern color=black}}
\addlegendentry{UL}


\draw[draw=black, fill=steelblue31119180,postaction={pattern=north east lines,pattern color=black}]
  (axis cs:-0.25, 0) rectangle (axis cs:-0.05, 1.0525);
\path[draw=black, thick] (axis cs:-0.15,1.01241)--(axis cs:-0.15,1.09259);

\draw[draw=black, fill=steelblue31119180,postaction={pattern=north east lines,pattern color=black}]
  (axis cs:0.75, 0) rectangle (axis cs:0.95, 1.2002);
\path[draw=black, thick] (axis cs:0.85,1.1807)--(axis cs:0.85,1.22033);

\draw[draw=black, fill=steelblue31119180,postaction={pattern=north east lines,pattern color=black}]
  (axis cs:1.75, 0) rectangle (axis cs:1.95, 1.2932);
\path[draw=black, thick] (axis cs:1.85,1.2400)--(axis cs:1.85,1.3463);

\draw[draw=black, fill=steelblue31119180,postaction={pattern=north east lines,pattern color=black}]
  (axis cs:2.75, 0) rectangle (axis cs:2.95, 1.1932);
\path[draw=black, thick] (axis cs:2.85,1.1400)--(axis cs:2.85,1.2463);

\draw[draw=black, fill=steelblue31119180,postaction={pattern=north east lines,pattern color=black}]
  (axis cs:3.75, 0) rectangle (axis cs:3.95, 1.6110);
\path[draw=black, thick] (axis cs:3.85,1.6210)--(axis cs:3.85,1.6010);

\draw[draw=black, fill=steelblue31119180,postaction={pattern=north west lines,pattern color=black}]
  (axis cs:4.75, 0) rectangle (axis cs:4.95, 0.6500);
\path[draw=black, thick] (axis cs:4.85,0.6300)--(axis cs:4.85,0.6700);

\end{axis}

\begin{axis}[
    width=0.8\textwidth,
    height=4.5cm,
    ybar,
    bar width=0pt,
    xmin=-0.5, xmax=5.5,
    ymin=0, ymax=175,
    ytick={0,25,50,75,100,125,150,175},
    ylabel={UL Throughput [Mbps]},
    ylabel style={color=darkorange25512714,font=\scriptsize},
    yticklabel style={color=darkorange25512714},
    axis y line*=right,
    axis x line=none,
    tick align=inside,
    ymajorgrids=false,
    xmajorgrids=false
]

\draw[draw=black, fill=darkorange25512714,postaction={pattern=north west lines,pattern color=black}]
  (axis cs:0.05,0) rectangle (axis cs:0.25,100.65);
\path[draw=black, thick] (axis cs:0.15,98.31)--(axis cs:0.15,103.00);

\draw[draw=black, fill=darkorange25512714,postaction={pattern=north west lines,pattern color=black}]
  (axis cs:1.05,0) rectangle (axis cs:1.25,100.80);
\path[draw=black, thick] (axis cs:1.15,98.48)--(axis cs:1.15,102.11);

\draw[draw=black, fill=darkorange25512714,postaction={pattern=north west lines,pattern color=black}]
  (axis cs:2.05,0) rectangle (axis cs:2.25,116.25);
\path[draw=black, thick] (axis cs:2.15,112.82)--(axis cs:2.15,119.68);

\draw[draw=black, fill=darkorange25512714,postaction={pattern=north west lines,pattern color=black}]
  (axis cs:3.05,0) rectangle (axis cs:3.25,106.25);
\path[draw=black, thick] (axis cs:3.15,102.82)--(axis cs:3.15,109.68);

\draw[draw=black, fill=darkorange25512714,postaction={pattern=north west lines,pattern color=black}]
  (axis cs:4.05,0) rectangle (axis cs:4.25,143.11);
\path[draw=black, thick] (axis cs:4.15,141.11)--(axis cs:4.15,145.11);

\draw[draw=black, fill=darkorange25512714,postaction={pattern=north west lines,pattern color=black}]
  (axis cs:5.05,0) rectangle (axis cs:5.25,123.11);
\path[draw=black, thick] (axis cs:5.15,121.11)--(axis cs:5.15,125.11);

\end{axis}
\end{tikzpicture}

\caption{Performance profiling of peak \gls{dl} and \gls{ul} throughput using iPerf. The configurations on the x-axis use the following nomenclature: $number\_of\_UEs-RU\_type-L1\_type$.}
\label{fig:2ue_static_rusim}

\end{figure}

Figure~\ref{fig:2ue_static_rusim} presents results from a selection of configurations, described on the x-axis using the following nomenclature: $number\_of\_UEs-RU\_type-L1\_type$. We include results for: 1 and 2 \glspl{ue}; \gls{ru} types including Foxconn (\textit{fxn}), Benetel (\textit{bntl}), VVDN (\textit{vvdn}), RuSIM/CoreSIM (\textit{sim}), and USRP \gls{sdr} N310 (\textit{sdr}); and Layer~1 implementations from NVIDIA (\textit{arc}) and \gls{oai} (\textit{oai}). For NVIDIA Aerial configuration, we use an NVIDIA Grace Hopper 200 server, while for the \gls{oai} Layer~1 configuration, we use an AMD EPYC9374F 32-core CPU.

The results leverage a DDDSU \gls{tdd} pattern (DDDDDDDSUU for \textit{fxn}) and a $100$~MHz bandwidth,  4~TRX antennas (2 for \textit{sdr}), 4~layers in the \gls{dl} (2 for \textit{sdr}), 1~layer in the \gls{ul}, and up to 256~QAM. The \gls{ota} measurements are done in an indoor laboratory environment with static \glspl{ue} at fixed locations while the frequency band varies depending on the capabilities of the \gls{ru}. In the simulated cases, measurements are done with a Keysight RuSIM device, which emulates the \gls{ru}, the wireless channel, and the \glspl{ue}, and Keysight CoreSIM to emulate the \gls{cn}~\cite{gemmi2024open6goticblueprintprogrammable}. We use the {\em ExcellentRadioConditions} channel model, which simulates a full-rank \gls{mimo} channel with high \gls{snr}.

Regarding the special slot, due to hardware limitations, it is not utilized in the \textit{fxn-arc} experiments, resulting in a lower \gls{dl} throughput compared to the other cases, particularly in the 1~\gls{ue} scenario (see~\cite{villa2024x5gtmc} for a detailed explanation). In the \textit{vvdn-oai}, \textit{bntl-oai}, and \textit{sdr-oai} configurations, we use 8~\gls{dl} and 2~\gls{ul} symbols, while in the \textit{sim-arc} case, 12~\gls{dl} and 1~\gls{ul} symbols are used.

In the emulated \textit{sim-arc} test cases, we achieve an average throughput of $1.61$~Gbps in the \gls{dl} and $143$~Mbps in the \gls{ul} with a single \gls{ue}, which is very close to the theoretical peak throughput. In the \gls{ota} tests, the average throughput drops to $1.3$~Gbps in the \gls{dl} and $116$~Mbps in the \gls{ul} for the case \textit{vvdn-oai} and to $1.2$~Gbps in the \gls{dl} and $100$~Mbps in the \gls{ul} for the case \textit{fxn-arc}.

Additionally, the RuSIM equipment supports an arbitrary number of \glspl{ue} for stress-testing the system. Figure~\ref{fig:25ues_rusim} shows the performance achieved as \glspl{ue} progressively attach, each pushing approximately $64$~Mbps, until reaching a total of 25 \glspl{ue} and $1.6$~Gbps, and then gradually disconnect.

This demonstrates how \gls{oai} can sustain a high number of simultaneously attached devices with fair resource sharing and no noticeable performance degradation. Moreover, other stress tests with this configuration have allowed up to 55~\glspl{ue} to remain attached at once, each running a \textit{ping} test to verify connectivity, further demonstrating the reliability of the network stack.

\begin{figure}
\centering
    \setlength\fwidth{0.9\linewidth}
    \setlength\fheight{.32\linewidth}
    \begin{tikzpicture}

\definecolor{lightgray204}{RGB}{204,204,204}
\definecolor{orange25512714}{RGB}{255,127,14}
\definecolor{steelblue31119180}{RGB}{31,119,180}

\begin{axis}[
    width=\fwidth,
    height=\fheight,
    legend style={
        at={(0.65,0.45)},
        fill opacity=1,
        draw opacity=1,
        text opacity=1,
        draw=black,
        font=\footnotesize
    },
    legend columns=1,
    xmajorgrids,
    x grid style={lightgray204},
    xlabel={Time [s]},
    xmin=0, xmax=110.08,
    xtick={0,20,40,60,80,100},
    xticklabels={0,20,40,60,80,100},
    xticklabel style={font=\footnotesize},
    ymin=0, ymax=1.750,
    ylabel={Throughput [Gbps]},
    ytick={0,0.350,0.700,1.050,1.400,1.750},
    ylabel style={color=steelblue31119180,font=\footnotesize},
    xlabel style={font=\footnotesize},
    ymajorgrids,
    y grid style={lightgray204},
    yticklabel style={color=steelblue31119180},
    axis y line*=left,
]

\addlegendimage{steelblue31119180, semithick, mark=*, mark size=1}
\addlegendentry{DL Throughput}
\addlegendimage{orange25512714, semithick, mark=x}
\addlegendentry{UE Count}

\addplot [
    semithick,
    steelblue31119180,
    mark=*,
    mark size=1,
    mark options={solid}
]
table {
x    y
0.0     0.0000196
1.28    0.0613451
2.56    0.0879709375
3.84    0.12583450625
5.12    0.1794034375
6.4     0.2067445625
7.68    0.2525313625
8.96    0.29029298125
10.24   0.317733925
11.52   0.37950935
12.8    0.4180307
14.08   0.4446775875
15.36   0.4988733875
16.64   0.53573201875
17.92   0.5708245125
19.2    0.62886126875
20.48   0.6531007125
21.76   0.69534871875
23.04   0.738907375
24.32   0.7726570125
25.6    0.82109881875
26.88   0.86415255
28.16   0.885225575
29.44   0.94418063125
30.72   0.9784839375
32.0    1.01082295625
33.28   1.07411298125
34.56   1.06954495625
35.84   1.10085355625
37.12   1.1595264125
38.4    1.2048861875
39.68   1.25131709375
40.96   1.30965701875
42.24   1.36094
43.52   1.34329131875
44.8    1.40832861875
46.08   1.439667975
47.36   1.47198305
48.64   1.51951234375
49.92   1.5713500625
51.2    1.5971448375
52.48   1.5742751875
53.76   1.5917046125
55.04   1.5648216875
56.32   1.59372583125
57.6    1.566130775
58.88   1.56438351875
60.16   1.5802058125
61.44   1.5169226125
62.72   1.4884696375
64.0    1.45230533125
65.28   1.3866329625
66.56   1.35571414375
67.84   1.317883975
69.12   1.2753097375
70.4    1.2486451
71.68   1.19470361875
72.96   1.15008650625
74.24   1.13239394375
75.52   1.07009030625
76.8    1.03429058125
78.08   1.00463335625
79.36   0.94398560625
80.64   0.9200554625
81.92   0.88167460625
83.2    0.83654514375
84.48   0.8003650375
85.76   0.7575958375
87.04   0.7107202
88.32   0.69031806875
89.6    0.630658875
90.88   0.59270321875
92.16   0.5648921125
93.44   0.50561626875
94.72   0.4812596375
96.0    0.44178194375
97.28   0.39110119375
98.56   0.354817725
99.84   0.31587101875
101.12  0.26352739375
102.4   0.24558925625
103.68  0.1890535375
104.96  0.1445029
106.24  0.12607159375
107.52  0.0636279375
108.8   0.03533138125
110.08  0.0000292125
};

\end{axis}

\begin{axis}[
    width=\fwidth,
    height=\fheight,
    x grid style={lightgray204},
    xmin=0, xmax=110.08,
    xtick={0,20,40,60,80,100},
    xticklabels={0,20,40,60,80,100},
    xticklabel style={font=\footnotesize},
    axis y line*=right,
    ylabel={Number of UEs},
    ylabel near ticks,
    ylabel style={color=orange25512714,font=\footnotesize},
    xlabel style={font=\footnotesize},
    ymin=0, ymax=25,
    ytick={0,5,10,15,20,25,25},
    yticklabel style={color=orange25512714},
]

\addplot [
    semithick,
    orange25512714,
    mark=x,
    mark size=2,
    mark options={solid}
]
table {
x     y
0.0      1
1.28     2
2.56     2
3.84     3
5.12     3
6.4      4
7.68     5
8.96     5
10.24    6
11.52    7
12.8     7
14.08    8
15.36    9
16.64    9
17.92    10
19.2     10
20.48    11
21.76    12
23.04    12
24.32    13
25.6     14
26.88    14
28.16    15
29.44    16
30.72    16
32.0     17
33.28    17
34.56    18
35.84    19
37.12    19
38.4     20
39.68    21
40.96    21
42.24    22
43.52    23
44.8     23
46.08    24
47.36    25
48.64    25
49.92    25
51.2     25
52.48    25
53.76    25
55.04    25
56.32    25
57.6     25
58.88    25
60.16    25
61.44    25
62.72    24
64.0     24
65.28    23
66.56    22
67.84    22
69.12    21
70.4     21
71.68    20
72.96    19
74.24    19
75.52    18
76.8     17
78.08    17
79.36    16
80.64    15
81.92    15
83.2     14
84.48    14
85.76    13
87.04    12
88.32    12
89.6     11
90.88    10
92.16    10
93.44    9
94.72    8
96.0     8
97.28    7
98.56    7
99.84    6
101.12   5
102.4    4
103.68   4
104.96   3
106.24   3
107.52   2
108.8    1
110.08   1
};

\end{axis}

\end{tikzpicture}

\caption{Performance profiling leveraging X5G with \gls{oai} NVIDIA Aerial and Keysight RuSIM/CoreSIM. The results show the downlink throughput as UEs progressively attach, each pushing approximately $64$~Mbps, until reaching a total of 25 and a peak of $1.6$~Gbps, and then gradually disconnect.}
\label{fig:25ues_rusim}

\end{figure}

When using an \gls{sdr} instead of an \gls{oru}, performance is typically lower. While it is theoretically possible on a \gls{usrp} N310 \gls{sdr} to drive four channels at $100$~MHz each, in practice it is challenging to sustain the required fronthaul throughput without incurring real-time faults. However, two channels at $100$~MHz each perform reliably, achieving $650$~Mbps in the \gls{dl} and $123$~Mbps in the \gls{ul}~\cite{OAI_jenkins}.

\begin{figure}
\centering
\includegraphics[width=0.99\linewidth]{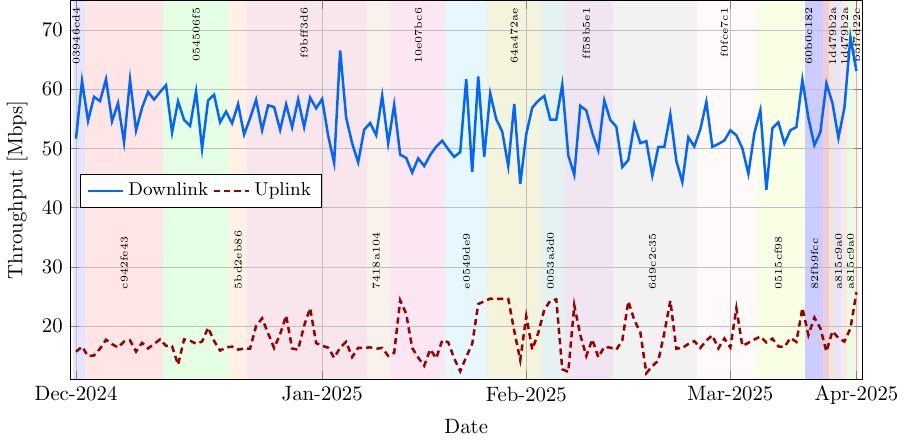}

\caption{History of the DL and UL throughput measured in the Colosseum pipeline. The different colors and the numbers show the commit IDs of the develop branch.}
\label{fig:colosseum_history}

\end{figure}

When using an \gls{sdr}- and \gls{oai}-based \gls{ue}, performance remains significantly lower than that of a \gls{cots} \gls{ue}. The best throughput measured so far is $70$~Mbps in the \gls{dl} and $25$~Mbps in the \gls{ul}. This has been achieved on the Colosseum testbed (see Section~\ref{sec:colosseum}), which also maintains a historical record of the pipeline’s results, which we plot in Figure~\ref{fig:colosseum_history}~\cite{openrangymwebsite}.

Additional performance results of \gls{oai} in various configurations can be found in the latest status reports of the different \gls{cicd} pipelines~\cite{OAI_jenkins}. However, it should be noted that these pipelines are not designed to maximize throughput, as their primary purpose is to test code stability rather than peak performance.

\section{End-to-end Systems Powered by OAI} 
\label{sec:end2end}

In this section, we provide a few examples of end-to-end testbeds that use \gls{oai} as well as other open-source projects. This list is by far not exhaustive but aims to give a few lighthouse examples.

\subsection{EURECOM Open5GLab}  
\label{sec:open5glab}

The EURECOM Open5GLab is a cutting-edge research and experimentation facility focused on open and virtualized 5G networks based on the OAI open-source software. It consists of computing and radio infrastructure interconnected through high-speed fiber links and telecom-grade switches. The computing infrastructure includes Multi-core Intel Xeon, AMD EPYC, as well as Grace Hopper 200 servers, which are used for \gls{ran}, \gls{cn}, and mobile-edge functions. Some servers also support inline acceleration (NVIDIA \gls{arc-ota}~\cite{arc-ota}) as well as look-aside acceleration (AMD T2 card).  Most of the computing resources are managed by Red Hat’s OpenShift container platform, but some nodes can be used as bare metal by experimenters and developers. The radio infrastructure includes \glspl{sdr} such as the USRP N310, X310, and B210, commercial split 8 \glspl{ru} from AW2S, and commercial split 7.2 O-RAN \glspl{ru} from Benetel, VVDN, and Foxconn. Some of these radios are deployed outdoors on EURECOM's roof where it has licenses to transmit in several 4G and 5G bands, such as Band 28 ($700$~MHz), Band~n38 ($2.6$~GHz TDD), Band~n78 ($3.5$~GHz TDD), Band~n77 ($3.8$~GHz TDD), Band~n258 ($25$~GHz TDD). Several \gls{cots} UEs, such as smartphones and modules, are also available in the lab. External access for onboarding software, collecting measurement data, and developing basic software for the site is available for partners using \gls{ssh} access. Some of these resources are also used by \gls{oai} Jenkins-based \gls{cicd} framework (see Section \ref{sec:cicd}).

The EURECOM Open5GLab is part of the SLICES project, which is a large-scale European initiative focused on creating a flexible research infrastructure for digital and network technologies~\cite{Fdida2022}. The deployment is documented in the SLICES 5G Blueprint, which allows other research sites to easily reproduce the setup~\cite{slices_5g_blueprint}.  

EURECOM also hosts the drone4wireless lab~\cite{drone4wireless}, which focuses on autonomous flying robots for sensing and connectivity. It features multiple drones equipped with 4G and 5G radios used for various experiments~\cite{Gangula2018,Mundlamuri2023}. 

The outdoor deployment has recently been extended with the Firecell GEO-5G testbed, which consists of 3 RUs with distributed antennas and which is used for 5G-based geolocalization~\cite{malik2024concept}.

\subsection{X5G} 
\label{sec:x5g}
X5G is a multi-vendor, O-RAN-compliant private \gls{5g} network deployed in the Boston, MA, campus of Northeastern University~\cite{villa2024x5gtmc}. It features the first-of-its-kind, 8-node deployment of the NVIDIA \gls{arc-ota} framework~\cite{arc-ota}, enabling the development and testing of next-generation wireless cellular networks on a programmable \gls{ota} platform with production-ready performance and capabilities.

As explained in Section~\ref{sec:ran}, the upper layers of the protocol stack, namely the \gls{cu} and \gls{du}-High, are deployed using the latest \gls{oai} software build, while the \gls{du}-Low utilizes the NVIDIA Aerial \gls{sdk}. \glspl{ru} from different vendors, including Foxconn \glspl{ru} operating in the n78 band and a Keysight \gls{ru} emulator, are integrated into the testbed. X5G incorporates \glspl{cn} from various projects, such as \gls{oai} and Open5GS~\cite{open5gs_website}. Finally, it includes the ``E'' release of the \gls{osc} \gls{near-rt-ric}, integrated through a custom E2 agent in \gls{oai}~\cite{moro2023nfv}, for the development of xApps.

All X5G software is containerized and deployed on a Red Hat OpenShift container platform cluster, enabling automation and orchestration of each component. The cluster comprises general-purpose hardware, including Dell, GIGABYTE, and Grace Hopper servers. Each GIGABYTE is equipped with an NVIDIA A100 GPU and Mellanox ConnectX-6 Dx \glspl{nic}, while each Grace Hopper features an NVIDIA Grace CPU Superchip, an NVIDIA H100 Tensor Core GPU, two BlueField-3 \glspl{dpu}, and ConnectX-7 \glspl{nic}. This configuration enables the execution of the NVIDIA Aerial GPU acceleration component. 

X5G supports the use of \gls{cots} \glspl{ue}, including devices from OnePlus, Apple, Samsung, and Pixel, as well as 5G modules such as Sierras and Quectel boards. Additionally, it supports the software \gls{ue} developed by \gls{oai} and described in Section~\ref{sec:soft-ue}.

\subsection{Colosseum} 
\label{sec:colosseum}

Colosseum is a publicly accessible testbed hosted at Northeastern University, part of the \gls{pawr} project~\cite{pawr}, as well as the world's largest wireless network emulator with hardware-in-the-loop~\cite{bonati2021colosseum}. This \gls{sdr}-based platform can be utilized as an Open \gls{ran} \gls{dt}~\cite{villa2023dt,polese2024colosseum}, enabling researchers to test and develop end-to-end solutions for wireless networks.

Thanks to its \gls{mchem}, consisting of 128~\glspl{usrp} X310 and a fabric of 64~FPGAs, Colosseum is capable of creating a \gls{dt} replica of a wide variety of real-world scenarios.

Additionally, with its 128 \glspl{srn}—each comprising a general-purpose server and an additional \gls{usrp} X310—users can remotely access the system via \gls{ssh}, deploy wireless protocol stacks, and run various experiments in a repeatable, reproducible, and controlled environment.

This capability is made possible by Colosseum's resource management framework, which facilitates the deployment of wireless solutions, such as \gls{oai}, from the \gls{cn} to the software \gls{ue}.

Moreover, its \gls{cicd} pipelines, described in Section~\ref{sec:cicd}, enable a \gls{ct} of the \gls{oai} software, for example, allowing the automatic testing of different releases across various \gls{dt} scenarios. Similarly, \gls{ct} pipelines in the \gls{ota} lab at Northeastern University enable the continuous testing of the \gls{oai} gNB with commercial smartphones and 5G modems~\cite{bonati20235gct}.

\begin{figure}
    \centering
    \includegraphics[width=0.99\linewidth]{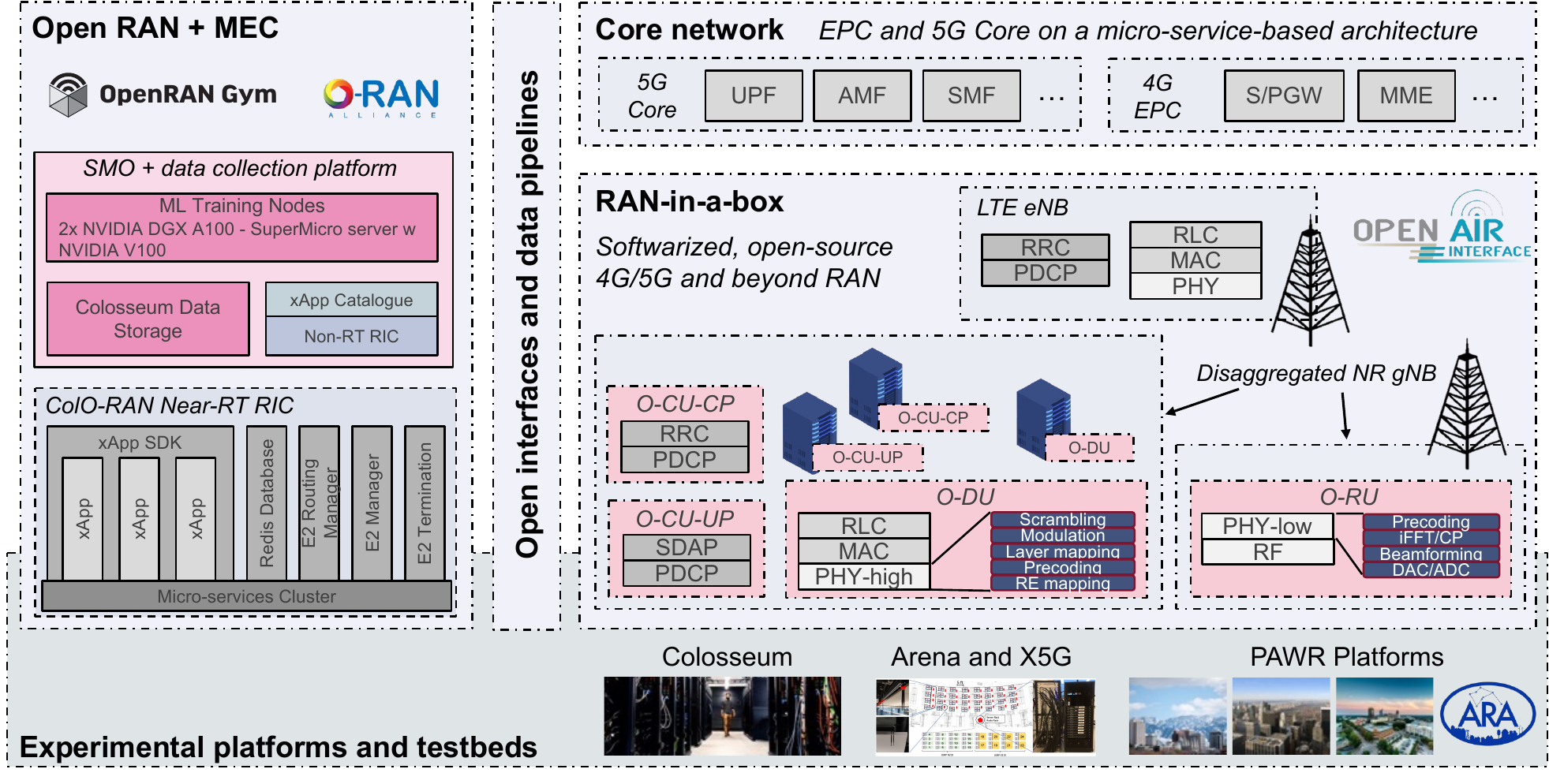}
    \caption{OAI and its role in the Open6G architecture deployed within Colosseum.}
    \label{fig:open6g}
\end{figure}

Within Colosseum, \gls{oai} is part of the Open6G blueprint, a collection of software solutions that enable research on open and programmable networks. The Open6G blueprint, also shown in Figure~\ref{fig:open6g}, combines open-source elements, interoperable interfaces, and research platforms to enable end-to-end studies on various O-RAN and next-generation wireless solutions. The same can be deployed in Colosseum or on \gls{ota} platforms such as X5G, discussed above.
OAI and this architecture have been featured in numerous Colosseum research projects and proof-of-concept solutions, both past and present. Tutorials on how to utilize \gls{oai} within the Colosseum platform are available on the OpenRAN Gym website~\cite{openrangymwebsite}, an open-source framework that promotes collaborative exploration and development in the O-RAN domain.

\subsection{POWDER} 

The POWDER platform, located in the University of Utah campus in Salt Lake City, UT, supports 5G research by offering advanced hardware and software tools, enabling seamless transition from simulations to real-world tests in both indoor and outdoor environments. It provides access to multiple 5G profiles, \gls{ota} testing setups, and controlled RF environments, with various \glspl{sdr} and 5G core networks like Open5GS and OAI. Researchers can create customized 5G networks, deploy RANs, and test novel applications such as spectrum sharing across networks. Outdoor experiments include mobile endpoints like campus shuttles equipped with 5G UEs~\cite{POWDER2024}.

\subsection{ARA Lab}

The ARA Wireless Living Lab at Iowa State University is a large-scale testbed advancing rural wireless technologies. It focuses on improving broadband and precision agriculture, supporting research in dynamic spectrum sharing, 5G/NextG, and low-latency communications.
The ARA-O-RAN testbed is a subset of the ARA Wireless Living Lab running OAI on custom-made BS based on USRP N320 together with \glspl{tmb} to amplify both transmitting and receiving signals and custom UEs built on USRP B210 with \glspl{ueb}, akin to a smaller version of the \glspl{tmb}. In their experiments, they were able to show a stable connection with UEs that are more than $1$~km away~\cite{boateng2024arasdr,zhang2024ara,OforiBoateng2025}.

\subsection{AERPAW}
AERPAW, or the Aerial Experimentation and Research Platform for Advanced Wireless, is a pioneering research platform hosted by North Carolina State University. It is focused on integrating \glspl{uav} into national airspace and advancing wireless technologies. The platform offers a software-defined, reproducible, and open-access environment for experimenting with 5G and beyond technologies. Similar to ARA, it features \glspl{gnb} and \glspl{ue} based on \gls{oai} and USRP \glspl{sdr} with the key difference that the \glspl{ue} can also be mounted on \glspl{uav}~\cite{Moore2024}.

\section{OAI Use-cases for 6G Research}	
\label{sec:6g}

While the official \textit{develop} branch of OAI only accepts standard-compliant contributions, there is significant work utilizing OAI for 6G research. In this section, we would like to highlight a few of the advanced research works that are currently outside the scope of the standard and thus are not on the official OAI roadmap.
However, it is conceivable that one day this research work will become part of the standard, which means that the code can be merged into the \textit{develop} branch. For example, this is what happened with the \gls{ntn} project, where the initial functionalities were developed along with standardization, and the first demonstrations were done using an initial specifications draft. Later, when the specifications were finalized in 3GPP Rel 17, the initial code was adapted and merged into the \textit{develop} branch of OAI.

\begin{table}
\caption{Summary of 6G challenges and use cases discussed in this section and list of related \gls{oai}-specific works.}
\begin{center}
\begin{tabular}{|c|c|}
\hline
{\bf 6G challenge} & {\bf OAI-related works} \\
\hline
\hline
Exploring new spectrum & \cite{Allbesmart2025FR3,Gangula2024,mundlamuri20245g} \\
\hline
Air Interface Enhancements & \cite{Cheng2024,Ramantas2023} \\
\hline
Architecture enhancements & \cite{Bonati2023,Boutiba2023,Lacava2025} \\
\hline
Energy Efficiency & \cite{Mohammadi2024,Shankaranarayanan2024,Martins2024,Centofanti2024} \\
\hline
Security and Resiliency & \cite{Kumar2022,Kumar2022a,Kumar2022b}\\
\hline
\end{tabular}
\end{center}
\label{tab:6gchallenges}
\end{table}

\subsection{Exploring New Spectrum}

Each new generation of mobile communication systems has introduced access to new spectrum bands, and 6G is no exception. The \gls{itu} has proposed a new frequency band, called the upper mid band or FR3, covering frequencies from $7.125$~GHz to $24.25$~GHz for 6G, expanding the currently defined FR1 ($410$~MHz~$-$~$7.125$~GHz) and FR2 ($24.25-71$~GHz) frequency ranges~\cite{ITU2024}. FR3 provides broader channels than FR1, while having lower propagation losses and cheaper hardware than FR2. It is more suitable for medium-coverage deployments and allows for efficient implementation of massive MIMO and 3D beamforming techniques~\cite{Bjoernson2024}.

Currently, \glspl{sdr} or \glspl{oru} for this frequency range are not widely available. One solution is to use external \gls{rf} frontends together with a standard \gls{sdr} such as the \gls{usrp}. One such frontend has been developed by the company PiRadio and demonstrated in cooperation with Allbesmart~\cite{Allbesmart2025FR3}. The demonstration makes use of the OAI UE, since the FR3 parameters are not yet standardized.  

Other challenges in the FR3 band are spectrum sharing since this frequency band is currently used by many other players. Although not done in FR3, OAI has been used to showcase such spectrum sharing capabilities~\cite{Gangula2024}.

Also, frequencies above the current limit of $71$~GHz have been proposed for 6G. In the so-called sub-terahertz frequency range, loosely defined as $71$~GHz to $300$~GHz, very large chunks of frequencies are available. While technically it is possible to leverage this spectrum (there has even been a demonstration of \gls{oai} at 130 GHz~\cite{mundlamuri20245g}), many problems still need to be solved, especially the management of ultra-sharp beams and mobility. 

\subsection{Air Interface Enhancements}

The 6G air interface will undergo significant advancements to enable ultra-high-speed, ultra-reliable, and intelligent communications. We expect the integration of AI/ML at different levels. Already in a 3GPP Rel.~18 study item, they explored the potential applications of AI and ML at the air interface level and their standardization~\cite{3GPP38.843}. The report identifies three key use cases: \gls{csi} feedback enhancement, beam management, and positioning accuracy enhancement.

\gls{csi} feedback enhancements are necessary to enable advanced multi-antenna processing such as cell-free MIMO, and massive MIMO.  \gls{csi} is typically estimated by the \gls{ue} and then fed back to the network in a compressed form. The accuracy of \gls{csi} directly influences the achievable spectral efficiency. Current \gls{csi} compression schemes are based on fixed rules and lack adaptability. However, deep learning-based \gls{csi} feedback has demonstrated superior performance compared to traditional methods~\cite{Guo2024}. One recent use case demonstration involving the \gls{oai} \gls{ue} showcases an \gls{ai}-enabled \gls{csi} feedback mechanism using an autoencoder on a real-time system ~\cite{Cheng2024}. This work demonstrates the capability of a programmable \gls{ue} to integrate \gls{ai} functions and validate them in real time. 

Another hot technology for the 6G air interface is \gls{ris}. These are artificial surfaces embedded with programmable reflective elements, often made of meta-materials, that can control the behavior of electromagnetic waves. By dynamically adjusting these elements, \gls{ris} can improve signal propagation and network performance in wireless environments. The integration of \gls{ris} into a real network is still an open research question. 

The 6G-BRICKS project addresses this challenge by integrating \gls{ris} into 5G \gls{ran}. This is achieved through the introduction of an \gls{ris} agent within the \gls{gnb} and an \gls{ris} control xApp, which is responsible for configuring and managing the \gls{ris}. The project aims to showcase this innovative technology using the \gls{oai} platform, laying the groundwork for practical deployment and experimentation~\cite{Ramantas2023}.

\subsection{Architecture Enhancements}

The O-RAN architecture with its open interfaces, \gls{ric} and xApps/rApps framework, as well as its data-driven approach, makes it the ideal environment for AI/ML integration. There are many examples where \gls{oai} is used to research and develop xApps and rApps using \gls{ai} and \gls{ml} techniques. In~\cite{Bonati2023}, the authors describe two xApps developed using OpenRAN Gym: a ``sched'' xApp that controls the scheduling policies for different classes of traffic, and a ``sched-slicing'' xApp that controls both the scheduling policies and the resource allocation (number of \glspl{rbg}) for each network slice. Another example is~\cite{Boutiba2023} where the authors use reinforcement learning to optimize \gls{tdd} patterns in the 5G system. 

xApps typically operate on the order of tens of milliseconds and have only access to certain data that is shared over the E2 interface.  However, some applications require even shorter latencies and also access to low-level data such as \gls{iq} samples. dApps were recently proposed as an extension of the O-RAN architecture. They are deployed directly on RAN nodes and can access data otherwise unavailable to RICs due to privacy or timing constraints, enabling the execution of control actions within shorter time intervals~\cite{Lacava2025}. The authors also present a prototype based on \gls{oai} and performance results in two real-time control use cases: spectrum sharing and positioning.  

\subsection{Energy Efficiency}

Energy efficiency has always been a design goal since the early days of mobile communication systems. In the context of modern 5G and future 6G networks, the focus increasingly shifts toward virtualized deployments. On one hand, virtualization can drastically reduce energy consumption in large-scale networks by pooling resources, dynamically reallocating them where needed, and placing unused cells into energy-saving modes. On the other hand, in smaller networks, virtualization may lead to higher overall energy consumption, as the infrastructure must be designed for peak load even when actual usage is low.

The use of hardware acceleration is one way of saving energy, and \gls{oai} supports both in-line and look-aside acceleration (see Section \ref{sec:ran}). Moreover, \gls{oai} can be used for experimentation with different energy-saving methods. For example, \cite{Mohammadi2024} uses \gls{oai} to experiment with management and orchestration frameworks that achieve energy efficiency improvements of 17.4\% per UE and 78.3\% per gNB. \cite{Shankaranarayanan2024} presents a similar study but adopts different methods for estimating energy consumption in the network. \cite{Martins2024}, on the other hand, uses the Linux \texttt{perf} tool to profile CPU cycle counts and cache miss ratios, leveraging these metrics as proxies for energy consumption. Finally, \cite{Centofanti2024} uses external power meters to measure the energy consumption of the radios only, specifically \gls{usrp} \glspl{sdr}, though the same approach can be applied to O-RUs.

\subsection{Security and Resiliency}

Just like energy efficiency, security has been a major design goal of mobile communication networks. While security leaks in 3GPP protocols have mostly been closed in 5G, current research focuses on security weaknesses in the new O-RAN interfaces. For example, \cite{Groen2024} studies the impact of encryption on E2 interface as well as the Open Fronthaul interface using the X5G testbed (see Section \ref{sec:x5g}). \cite{Wen2024} on the other hand develops an xApp employing unsupervised anomaly detection and LLM-based expert referencing to detect and analyze emerging threats and anomalies at run-time. They  demonstrate a prototype of their system on an O-RAN compliant cellular network testbed based on \gls{oai}. 

\subsection{Ubiquitous Connectivity}

Ubiquitous connectivity in 6G refers to seamless, high-speed, and reliable network access available everywhere, anytime, for anyone and anything. One approach to achieving this is through the use of \gls{ntn}, based on satellites or high-altitude platforms, to deliver 5G wireless communication services. Support for \gls{ntn} with both \gls{geo} and \gls{leo} transparent satellites has been integrated into OAI and demonstrated in a series of papers~\cite{Kumar2022,Kumar2022a,Kumar2022b}. In 6G, a tighter integration between terrestrial and non-terrestrial networks is expected.

\section{Conclusions and Future Directions}
\label{sec:conclusions}

This paper explored the \gls{oai} project, an open-source initiative that plays a pivotal role in the evolution of 5G and the future 6G networks. We presented the history and progression of \gls{oai}, its unique licensing model, and the various projects under its umbrella. We also highlighted the development methodology, \gls{cicd} processes, and end-to-end systems powered by \gls{oai}. The open-source approach of \gls{oai} is emphasized as essential for tackling the challenges of 6G, fostering community collaboration, and driving innovation in next-generation wireless technologies.

While \gls{ai} and \gls{ml} have come only as an afterthought within the already established 5G architecture and air interface, 6G will be the first standard to embrace \gls{ai} natively~\cite{lin2023embracing}.

Our vision for \gls{oai} is to establish it as the definitive reference implementation for 6G as the standard takes shape. Achieving this requires not only robust and reliable implementations of both the \gls{ran} and \gls{ue} but also a forward-thinking architecture to support the integration of cutting-edge AI/ML algorithms. Seamless collaboration between \gls{ran} and \glspl{ue} is critical to enabling researchers and developers to easily deploy and test their models. By leveraging frameworks such as O-RAN xApps on the \gls{ran} side and extending similar plug-and-play capabilities to the \gls{ue}, OAI can foster innovation across all layers of the network. This will enable researchers to focus on groundbreaking advancements without being hindered by system integration challenges, paving the way for the development of 6G.

\section*{Acknowledgements}
As described in the paper, OAI is a community-driven project, and the authors of this paper would like to acknowledge and thank every single one of its contributors. It would be too much to list them here, but the interested reader can go to GitLab's analytics page~\cite{oai_git} and browse the names of the different contributors. There are also countless people behind the scenes, most notably the \gls{osa} team~\cite{osateam} as well as the \gls{osa} board of directors~\cite{osamembers}. 

\section*{Funding Sources}
The authors from Northeastern University were partially supported by Open6G project through the Office of the Under Secretary of Defense for Research and Engineering (OUSD(R\&E)) under Army Research Laboratory Cooperative Agreement Number W911NF-24-2-0065, the National Telecommunications and Information Administration (NTIA)’s Public Wireless Supply Chain Innovation Fund (PWSCIF) under Awards No. 25-60-IF002, 25-60-IF011, and 25-60-IF054, as well as the x-mili project through the U.S. National Science Foundation under grant CNS-2117814.  

\section*{Declaration}
During the preparation of this work the author(s) used generative AI tools such as chatGPT in order to improve the language of the paper. After using this tool/service, the author(s) reviewed and edited the content as needed and take(s) full responsibility for the content of the published article.

\bibliographystyle{elsarticle-num} 
\bibliography{references.bib,biblio.bib}

\begin{thebibliography}{100}
\expandafter\ifx\csname url\endcsname\relax
  \def\url#1{\texttt{#1}}\fi
\expandafter\ifx\csname urlprefix\endcsname\relax\def\urlprefix{URL }\fi
\expandafter\ifx\csname href\endcsname\relax
  \def\href#1#2{#2} \def\path#1{#1}\fi

\bibitem{Larsson2024}
D.~C. Larsson, A.~Grövlen, S.~Parkvall, O.~Liberg,
  \href{https://www.ericsson.com/en/blog/2024/3/6g-standardization-timeline-and-technology-principles}{{6G}
  standardization – an overview of timeline and high-level technology
  principles}, Ericsson Blog (Mar. 2024).
\newline\urlprefix\url{https://www.ericsson.com/en/blog/2024/3/6g-standardization-timeline-and-technology-principles}

\bibitem{ITU-R-M.2160-0}
{International Telecommunication Union}, {Framework and overall objectives of
  the future development of {IMT} for 2030 and beyond}, Recommendation ITU-R
  M.2160-0 (Nov. 2023).

\bibitem{ngmn}
{Next Generation Mobile Networks Alliance},
  \href{https://www.ngmn.org/wp-content/uploads/NGMN_6G_Position_Statement.pdf}{{6G}
  position statement - an operator view} (Sep. 2023).
\newline\urlprefix\url{https://www.ngmn.org/wp-content/uploads/NGMN_6G_Position_Statement.pdf}

\bibitem{O-RAN6G}
V.~Dixit, J.~Plachy, K.~Sun, A.~Ikami, E.~O.~K. Lee,
  \href{https://mediastorage.o-ran.org/ngrg-rr/nGRG-RR-2023-01-O-RAN-Towards-6G-v1_3.pdf}{{O-RAN}
  towards {6G}}, resreport RR-2023-01, {O-RAN} next Generation Research Group
  ({nGRG}) (2023).
\newline\urlprefix\url{https://mediastorage.o-ran.org/ngrg-rr/nGRG-RR-2023-01-O-RAN-Towards-6G-v1_3.pdf}

\bibitem{Kerboeuf2024}
S.~Kerboeuf, P.~Porambage, A.~Jain, P.~Rugeland, G.~Wikström, M.~Ericson,
  D.~Thai~Bui, A.~Outtagarts, H.~Karvonen, P.~Alemany, R.~Muñoz, R.~Vilalta,
  P.~Botsinis, A.~Ramos, J.~Castaneda~Cisneros, M.~Karaca, C.~Karousatou,
  S.~Barmpounakis, P.~Demestichas, A.~Zafeiropoulos, I.~Tzanettis,
  S.~Papavassiliou, P.~G. Giardina, G.~Landi, B.~Han, A.~Nimr, M.~A. Uusitalo,
  {Design Methodology for 6G End-to-End System: Hexa-X-II Perspective}, IEEE
  Open Journal of the Communications Society 5 (2024) 3368--3394.
\newblock \href {https://doi.org/10.1109/OJCOMS.2024.3398504}
  {\path{doi:10.1109/OJCOMS.2024.3398504}}.

\bibitem{o-ran-webpage}
{O-RAN Alliance} (Nov. 2024).
\newblock \href{https://www.o-ran.org/}{[link]}.
\newline\urlprefix\url{https://www.o-ran.org/}

\bibitem{Bhat2021}
A.~Bhat, N.~Gupta, J.~Thaliath, R.~Banerji, V.~Sapru, S.~Singh, {Role of
  Open-Source in 6G Wireless Networks}, in: 6G Mobile Wireless Networks,
  Springer, 2021, pp. 379--392.

\bibitem{BONATI2020107516}
L.~Bonati, M.~Polese, S.~D’Oro, S.~Basagni, T.~Melodia,
  \href{https://www.sciencedirect.com/science/article/pii/S1389128620311786}{{Open,
  Programmable, and Virtualized {5G} Networks: State-of-the-Art and the Road
  Ahead}}, Computer Networks 182 (2020) 107516.
\newblock \href {https://doi.org/https://doi.org/10.1016/j.comnet.2020.107516}
  {\path{doi:https://doi.org/10.1016/j.comnet.2020.107516}}.
\newline\urlprefix\url{https://www.sciencedirect.com/science/article/pii/S1389128620311786}

\bibitem{kaltenberger2020}
F.~Kaltenberger, A.~P. Silva, A.~Gosain, L.~Wang, T.-T. Nguyen,
  \href{https://www.sciencedirect.com/science/article/pii/S1389128619314410}{{OpenAirInterface:
  Democratizing innovation in the {5G} Era}}, Computer Networks 176 (2020)
  107284.
\newblock \href {https://doi.org/https://doi.org/10.1016/j.comnet.2020.107284}
  {\path{doi:https://doi.org/10.1016/j.comnet.2020.107284}}.
\newline\urlprefix\url{https://www.sciencedirect.com/science/article/pii/S1389128619314410}

\bibitem{Zayen2015}
B.~Zayen, F.~Kaltenberger, R.~Knopp, Opportunistic Spectrum Sharing and White
  Space Access: The Practical Reality, Wiley, 2015, Ch. OpenAirInterface and
  ExpressMIMO2 for spectrally agile communication.

\bibitem{usrp}
{ETTUS Research}, \href{https://www.ettus.com/}{{The universal software radio
  peripheral {USRP} Software Defined Radio Device}} (Nov. 2024).
\newline\urlprefix\url{https://www.ettus.com/}

\bibitem{Nikaein2014}
N.~{N}ikaein, R.~{K}nopp, F.~{K}altenberger, L.~{G}authier, C.~{B}onnet,
  D.~{N}ussbaum, R.~{G}haddab,
  \href{http://www.eurecom.fr/publication/4371}{{O}pen{A}ir{I}nterface 4{G}: an
  open {LTE} network in a {PC}}, in: Proceedings of the 20th annual
  international conference on Mobile computing and networking (MobiCom '14),
  {M}aui, 2014.
\newline\urlprefix\url{http://www.eurecom.fr/publication/4371}

\bibitem{osa_board_members}
{OpenAirInterface Software Alliance},
  \href{https://openairinterface.org/osa-board-members/}{{Our people}}, online
  (2024).
\newline\urlprefix\url{https://openairinterface.org/osa-board-members/}

\bibitem{SCF2021FAPI}
{Small Cell Forum}, {5G} {FAPI}: {PHY} {API} {S}pecification, techreport
  222.10.04 (November 2021).

\bibitem{OSA-ORAN-MoU}
{OpenAirInterface Software Alliance},
  \href{https://openairinterface.org/news/o-ran-alliance-and-openairinterface-software-alliance-expand-cooperation-on-developing-open-software-for-the-ran/}{{O-RAN
  ALLIANCE} and {OpenAirInterface} software alliance expand cooperation on
  developing open software for the {RAN}} (2024).
\newline\urlprefix\url{https://openairinterface.org/news/o-ran-alliance-and-openairinterface-software-alliance-expand-cooperation-on-developing-open-software-for-the-ran/}

\bibitem{OSA-LF-MoU}
{OpenAirInterface Software Alliance},
  \href{https://openairinterface.org/news/lf-networking-and-openairinterface-software-alliance-sign-mou-with-openairinterface/}{{LF}
  networking and openairinterface software alliance sign {MoU}} (2024).
\newline\urlprefix\url{https://openairinterface.org/news/lf-networking-and-openairinterface-software-alliance-sign-mou-with-openairinterface/}

\bibitem{scf}
{Small Cell Forum}.
\newblock \href{https://www.smallcellforum.org/}{[link]}.
\newline\urlprefix\url{https://www.smallcellforum.org/}

\bibitem{osi-website}
{The Open Source Initiative}, \href{http://www.opensource.org}{{website}} (Nov.
  2024).
\newline\urlprefix\url{http://www.opensource.org}

\bibitem{gnu-gpl-v2}
{Free Software Foundation},
  \href{https://www.gnu.org/licenses/old-licenses/gpl-2.0.en.html}{{GNU}
  general public license, version 2} (2024).
\newline\urlprefix\url{https://www.gnu.org/licenses/old-licenses/gpl-2.0.en.html}

\bibitem{oai-public-license}
{OpenAirInterface Software Alliance},
  \href{https://openairinterface.org/legal/oai-public-license/}{{OAI} public
  license v1.1} (2024).
\newline\urlprefix\url{https://openairinterface.org/legal/oai-public-license/}

\bibitem{ETSI2024}
{European Telecommunications Standards Institute},
  \href{https://ipr.etsi.org/}{{IPR} online database} (2024).
\newline\urlprefix\url{https://ipr.etsi.org/}

\bibitem{srsran}
{Software Radio Systems}, \href{https://www.srsran.com/}{{The {srsRAN}
  project}}.
\newline\urlprefix\url{https://www.srsran.com/}

\bibitem{Amini2024}
M.~Amini, C.~Rosenberg, \href{https://arxiv.org/abs/2407.02341}{{Performance
  Analysis and Comparison of Full-Fledged 5G Standalone Experimental {TDD}
  Testbeds in Single \& Multi-{UE} Scenarios}} (2024).
\newblock \href {http://arxiv.org/abs/2407.02341} {\path{arXiv:2407.02341}}.
\newline\urlprefix\url{https://arxiv.org/abs/2407.02341}

\bibitem{Alves2024}
R.~P. Alves, J.~G.~A. da~S.~Alves, M.~R. Camelo, W.~O. de~Feitosa, V.~F.
  Monteiro, F.~R.~P. Cavalcanti,
  \href{https://arxiv.org/abs/2406.01485}{{Experimental comparison of {5G}
  {SDR} platforms: {srsRAN} x {OpenAirInterface}}} (2024).
\newblock \href {http://arxiv.org/abs/2406.01485} {\path{arXiv:2406.01485}}.
\newline\urlprefix\url{https://arxiv.org/abs/2406.01485}

\bibitem{free5gc}
{Linux Foundation}, \href{https://free5gc.org/}{{Open Source 5G Core Network
  Implementation}}.
\newline\urlprefix\url{https://free5gc.org/}

\bibitem{open5gs}
\href{https://open5gs.org/}{{Open Source implementation for {5G} Core and
  {EPC}}}.
\newline\urlprefix\url{https://open5gs.org/}

\bibitem{OSAprojects}
{OpenAirInterface Software Alliance},
  \href{https://openairinterface.org/projects/}{{OpenAirInterface} projects}
  (2024).
\newline\urlprefix\url{https://openairinterface.org/projects/}

\bibitem{OSA23fronthaul}
{OpenAirInterface Software Alliance}, {End-to-End Over-the-Air {5G} Demo with
  {OAI} {CU}-{DU} and {LITEON} {O-RU}}, O-RAN F2F Meeting, June 2023, Osaka,
  Japan (Jun. 2023).

\bibitem{villa2024x5g}
D.~Villa, I.~Khan, F.~Kaltenberger, N.~Hedberg, R.~S. da~Silva, A.~Kelkar,
  C.~Dick, S.~Basagni, J.~M. Jornet, T.~Melodia, M.~Polese, D.~Koutsonikolas,
  {An Open, Programmable, Multi-vendor 5G O-RAN Testbed with NVIDIA ARC and
  OpenAirInterface}, in: Proc. of the 2nd Workshop on Next-generation Open and
  Programmable Radio Access Networks (NG-OPERA), 2024.

\bibitem{OSA23ARC}
{OpenAirInterface Software Alliance}, {NVIDIA} {ARC} with {OAI} {L2+} for next
  generation innovation, O-RAN F2F Meeting, June 2023, Osaka, Japan (Jun.
  2023).

\bibitem{OSA23lookaside}
{OpenAirInterface Software Alliance}, {End-to-End {5G} Demo featuring Lookaside
  Acceleration of {LDPC} Encode/Decode and {O-RAN} {F1} and 7.2 Splits}, O-RAN
  F2F Meeting, October 2023, Phoenix, AZ, USA (Oct. 2023).

\bibitem{OSA24lookaside}
{OpenAirInterface Software Alliance}, {Showcasing Fronthaul 7.2 and Lookaside
  Hardware Acceleration of {5G}}, O-RAN F2F Meeting, February 2024, Athens,
  Greece (Feb. 2024).

\bibitem{Schmidt2023}
R.~Schmidt, M.~Irazabal, N.~Nikaein, {Improving {QoE} using {OSC} near-{RT}
  {RIC} and {OAI} {5G} {RAN} leveraging {O-RAN} {E2} {KPM} and {RC} {SMs}},
  O-RAN F2F meeting, Phoenix (AZ), United States (Oct. 2023).

\bibitem{OSA23OSC-RIC}
{OpenAirInterface Software Alliance}, {Improving {QoE} using {OSC} near-{RT}
  {RIC} and {OAI} {5G} {RAN} leveraging {O-RAN} {E2SM}-{KPM} and {E2SM}-{RC}},
  O-RAN F2F Meeting, October 2023, Phoenix, AZ, USA (Oct. 2023).

\bibitem{schmidt21flexRIC}
R.~Schmidt, M.~Irazabal, N.~Nikaein,
  \href{https://doi.org/10.1145/3485983.3494870}{{FlexRIC}: an {SDK} for
  next-generation {SD}-{RANs}}, in: Proceedings of the 17th International
  Conference on Emerging Networking EXperiments and Technologies, CoNEXT '21,
  Association for Computing Machinery, New York, NY, USA, 2021, p. 411–425.
\newblock \href {https://doi.org/10.1145/3485983.3494870}
  {\path{doi:10.1145/3485983.3494870}}.
\newline\urlprefix\url{https://doi.org/10.1145/3485983.3494870}

\bibitem{OSA23FlexRIC}
{OpenAirInterface Software Alliance}, {Improving {QoE} using {O-RAN} compliant
  near-{RT} {RIC}, {KPM} v03.00 and {RC} v01.03 {SMs} in an online multi-player
  game}, O-RAN F2F Meeting, June 2023, Osaka, Japan (Jun. 2023).

\bibitem{Kasparik2023}
M.~Kasparik, {Non-real-time Control of OpenAirInterface using the {O-RAN}
  {O1}-interface}, Joint OSC/OSFG-OAI Workshop: End-to-End Reference Designs
  for O-RAN, Boston (MA), United States (Nov. 2023).

\bibitem{OSA23O1}
{OpenAirInterface Software Alliance}, {5G} {RAN} management through an
  {SMO}/non-{RT} {RIC} via the {O1} interface, O-RAN F2F Meeting, October 2023,
  Phoenix, AZ, USA (Oct. 2023).

\bibitem{open5gs_website}
{Open5GS}, \href{https://open5gs.org}{{Open5GS} website}, accessed August 2024
  (2024).
\newline\urlprefix\url{https://open5gs.org}

\bibitem{free5gc_website}
{free5GC}, \href{https://free5gc.org/}{{free5GC} website} (2024).
\newline\urlprefix\url{https://free5gc.org/}

\bibitem{oai_git}
{OpenAirInterface Software Alliance},
  \href{https://gitlab.eurecom.fr/oai/openairinterface5g/}{{OpenAirInterface}
  git repository} (2024).
\newline\urlprefix\url{https://gitlab.eurecom.fr/oai/openairinterface5g/}

\bibitem{Kumar2022}
S.~Kumar, A.~Abdalla, O.~Kodheli, J.~Querol, S.~Chatzinotas, T.~Schlichter,
  G.~Casati, T.~Heyn, F.~Volk, S.~Kaya, et~al., {5G}-{NTN} {GEO}-based in-lab
  demonstrator using openairinterface {5G}, in: 11th Advanced Satellite
  Multimedia Conference, 2022.

\bibitem{Kumar2022a}
S.~Kumar, A.~K. Meshram, A.~Astro, J.~Querol, T.~Schlichter, G.~Casati,
  T.~Heyn, F.~V{\"o}lk, R.~T. Schwarz, A.~Knopp, et~al., {OpenAirInterface as a
  platform for {5G}-{NTN} Research and Experimentation}, in: 2022 IEEE Future
  Networks World Forum (FNWF), IEEE, 2022, pp. 500--506.

\bibitem{Kumar2022b}
S.~Kumar, O.~Kodheli, A.~Astro, J.~Querol, S.~Chatzinotas, G.~Casati,
  T.~Schlichter, T.~Heyn, H.~Cheporniuk, F.~V{\"o}lk, et~al., {5G}-{NTN}
  {GEO}-based over-the-air demonstrator using openairinterface, in: 39th
  International Communications Satellite Systems Conference (ICSSC 2022), Vol.
  2022, IET, 2022, pp. 110--114.

\bibitem{TMYTEK2024}
TMYTEK, \href{https://tmytek.com/solutions/mmW-OAI}{{mmW-OAI}} (Jul. 2024).
\newline\urlprefix\url{https://tmytek.com/solutions/mmW-OAI}

\bibitem{Cheng2024a}
H.~Cheng, S.~D'Oro, R.~Gangula, S.~Velumani, D.~Villa, L.~Bonati, M.~Polese,
  G.~Arrobo, C.~Maciocco, T.~Melodia,
  \href{https://arxiv.org/abs/2410.12978}{{{ORANSlice}: An Open-Source {5G}
  Network Slicing Platform for {O-RAN}}}, in: {Open-AI RAN ’24}: Proceedings
  of the 1st {ACM} Workshop on Open and {AI RAN}, Washington, D.C., USA, 2024.
\newblock \href {http://arxiv.org/abs/2410.12978} {\path{arXiv:2410.12978}}.
\newline\urlprefix\url{https://arxiv.org/abs/2410.12978}

\bibitem{5G-OPERA_D52}
{5G-OPERA Project Consortium},
  \href{https://franco-german-5g-ecosystem.eu/deliverables-or-reports/}{{Time-Sensitive
  Networks, {QoS}}}, Deliverable D5.2 (Dec. 2024).
\newline\urlprefix\url{https://franco-german-5g-ecosystem.eu/deliverables-or-reports/}

\bibitem{malik2024concept}
A.~Malik, M.~Ahadi, F.~Kaltenberger, K.~Warnke, N.~T. Thinh, N.~Bouknana,
  C.~Thienot, G.~Onche, S.~Arora, {From Concept to Reality: {5G} Positioning
  with Open-Source Implementation of {UL}-{TDoA} in OpenAirInterface}, arXiv
  preprint arXiv:2409.05217 (2024).

\bibitem{Elkadi2023}
M.~Elkadi, D.~Kim, E.~Ahmed, A.~Le, M.~Sadeghi, P.~Russell, B.~Ryu, {Open
  Source-Based Over-the-Air {5G} New Radio Sidelink Testbed}, in: MILCOM 2023 -
  2023 IEEE Military Communications Conference (MILCOM), 2023, pp. 63--70.
\newblock \href {https://doi.org/10.1109/MILCOM58377.2023.10356257}
  {\path{doi:10.1109/MILCOM58377.2023.10356257}}.

\bibitem{Mehlhose2024Sidelink}
M.~Mehlhose, {An open source {6G} testbed for {FR2} Sidelink as well as Joint
  Communication and Sensing research}, OpenAirInterface 10th Anniversary
  Workshop (Sep. 2024).

\bibitem{irazabal2023tcRAN}
M.~Irazabal, N.~Nikaein,
  \href{https://doi.org/10.1109/JSAC.2023.3336162}{{TC-RAN}: A programmable
  traffic control service model for {5G}/{6G} {SD-RAN}}, IEEE J.Sel. A. Commun.
  42~(2) (2023) 406–419.
\newblock \href {https://doi.org/10.1109/JSAC.2023.3336162}
  {\path{doi:10.1109/JSAC.2023.3336162}}.
\newline\urlprefix\url{https://doi.org/10.1109/JSAC.2023.3336162}

\bibitem{foukas23janus}
X.~Foukas, B.~Radunovic, M.~Balkwill, Z.~Lai, C.~Settle,
  \href{https://doi.org/10.1145/3570361.3614065}{{Programmable {RAN} Platform
  for Flexible Real-Time Control and Telemetry}}, in: Proceedings of the 29th
  Annual International Conference on Mobile Computing and Networking, ACM
  MobiCom '23, Association for Computing Machinery, New York, NY, USA, 2023.
\newblock \href {https://doi.org/10.1145/3570361.3614065}
  {\path{doi:10.1145/3570361.3614065}}.
\newline\urlprefix\url{https://doi.org/10.1145/3570361.3614065}

\bibitem{Ko2024edgeric}
W.-H. Ko, U.~Ghosh, U.~Dinesha, R.~Wu, S.~Shakkottai, D.~Bharadia,
  \href{https://www.usenix.org/conference/nsdi24/presentation/ko}{{EdgeRIC}:
  Empowering real-time intelligent optimization and control in {NextG} cellular
  networks}, in: 21st USENIX Symposium on Networked Systems Design and
  Implementation (NSDI 24), USENIX Association, Santa Clara, CA, 2024, pp.
  1315--1330.
\newline\urlprefix\url{https://www.usenix.org/conference/nsdi24/presentation/ko}

\bibitem{Vladic2023}
T.~Vladic, {RAN} control of {UE}-specific {DRB}-level {QoS} attributes, Joint
  OSC/OSFG-OAI Workshop: End-to-End Reference Designs for O-RAN, Boston (MA),
  United States (Nov. 2023).

\bibitem{srsRAN2024}
{srsRAN},
  \href{https://docs.srsran.com/projects/project/en/latest/tutorials/source/near-rt-ric/source/index.html}{{O-RAN
  NearRT-RIC and xApp}}, online (Nov. 2024) [cited 05.11.2024].
\newline\urlprefix\url{https://docs.srsran.com/projects/project/en/latest/tutorials/source/near-rt-ric/source/index.html}

\bibitem{Feraudo2024}
A.~Feraudo, S.~Maxenti, A.~Lacava, P.~Bellavista, M.~Polese, T.~Melodia,
  {xDevSM: Streamlining xApp Development With a Flexible Framework for O-RAN E2
  Service Models}, in: Proceedings of ACM WiNTECH, Washington, DC, USA, 2024.

\bibitem{Yonan2024}
M.~Yonan, {RIC}-{TaaP}: {RIC} testing as a platform, OpenAirInterface 10th
  Anniversary Workshop , Biot , France (Sep. 2024).

\bibitem{oaifeatures}
{OpenAirInterface Software Alliance},
  \href{https://gitlab.eurecom.fr/oai/openairinterface5g/-/blob/develop/doc/FEATURE_SET.md}{{OpenAirInterface}
  feature set} (Nov. 2024).
\newline\urlprefix\url{https://gitlab.eurecom.fr/oai/openairinterface5g/-/blob/develop/doc/FEATURE_SET.md}

\bibitem{Mongazon2024}
B.~Mongazon, L.~Thomas, {Compliance of {OAI} Soft {UE} to {Nokia} {gNB}},
  OpenAirInterface 10th anniversary workshop (2024).

\bibitem{3gpp38.213}
{3GPP}, {Physical layer procedures for control}, {Technical Specification}
  38.213 v18.4 (Sep. 2024).

\bibitem{3gpp38.214}
{3GPP}, {Physical layer procedures for data}, {Technical Specification} 38.214
  v18.4 (Sep. 2024).

\bibitem{Kumar2023}
S.~Kumar, C.~K. Sheemar, J.~Querol, T.~Yilmaz, S.~Chatzinotas, M.~Hammouda,
  T.~Heyn, T.~Schlichter, P.~Marques, L.~Pereira, R.~Magueta, A.~Kapovits,
  S.~Cioni, M.~Conti, {5G} {NTN} {LEO} based demonstrator using
  openairinterface {5G}, in: 40th International Communications Satellite
  Systems Conference (ICSSC 2023), Vol. 2023, 2023, pp. 69--75.
\newblock \href {https://doi.org/10.1049/icp.2024.0822}
  {\path{doi:10.1049/icp.2024.0822}}.

\bibitem{PeralRosado2024}
J.~A. del Peral-Rosado, A.~Y. Yildirim, A.~Soderini, R.~Mundlamuri,
  F.~Kaltenberger, E.~Rastorgueva-Foi, J.~Talvitie, I.~Lapin, D.~Flachs,
  {Initial Experimentation of a Real-Time {5G} mmWave Downlink Positioning
  Testbed}, in: Engineering Proceedings, 2024.

\bibitem{PeralRosado2024a}
J.~A. del Peral-Rosado, S.~Schlötzer, E.~Ince, P.~Nolle, F.~Kaltenberger,
  N.~Sirola, S.~Garlaschi, L.~Canzian, I.~Lapin, D.~Flachs, {Sub-Meter Hybrid
  Positioning with Flying {5G} Networks and Synchronization Corrections}, in:
  ION GNSS+, Technical meeting and showcase of GNSS technology, products and
  services, Baltimore, MD, 2024.

\bibitem{Palama2023}
I.~Palam{\`a}, S.~Bartoletti, G.~Bianchi, N.~Blefari~Melazzi, {Experimental
  assessment of {SDR}-based {5G} positioning: methodologies and insights},
  Annals of Telecommunications (2023) 1--13.

\bibitem{Mundlamuri2024}
R.~Mundlamuri, R.~Gangula, F.~Kaltenberger, R.~Knopp, {Novel Round Trip Time
  Estimation in {5G} {NR}}, in: 2024 IEEE Global Communications Conference:
  Wireless Communications (Globecom 2024 WC), Cape Town, South Africa, 2024,
  available as arXiv preprint arXiv:2404.19618.

\bibitem{etsi_osm}
ETSI, \href{https://osm.etsi.org/}{{Open Source MANO}} (Dec. 2024).
\newline\urlprefix\url{https://osm.etsi.org/}

\bibitem{onap}
{Open Network Automation Platform (ONAP)}, \href{https://www.onap.org/}{{ONAP
  Project}} (Nov. 2024).
\newline\urlprefix\url{https://www.onap.org/}

\bibitem{sylva}
{Sylva Community}, \href{https://sylvaproject.org/}{{Sylva Project}} (Nov.
  2024).
\newline\urlprefix\url{https://sylvaproject.org/}

\bibitem{nephio}
{Nephio Community}, \href{https://nephio.org/}{{Nephio Project}} (Nov. 2024).
\newline\urlprefix\url{https://nephio.org/}

\bibitem{OSAtestbenches}
{OpenAirInterface Software Alliance},
  \href{https://gitlab.eurecom.fr/oai/openairinterface5g/-/blob/develop/doc/TESTBenches.md}{{OpenAirInterface}
  {CI/CD} testbenches} (Nov. 2024).
\newline\urlprefix\url{https://gitlab.eurecom.fr/oai/openairinterface5g/-/blob/develop/doc/TESTBenches.md}

\bibitem{openrangymwebsite}
{OpenRAN Gym}, \href{https://openrangym.com/}{{OpenRAN Gym Website}}, accessed
  August 2024 (2024).
\newline\urlprefix\url{https://openrangym.com/}

\bibitem{osa-dockerhub}
{OpenAirInterface Software Alliance},
  \href{https://hub.docker.com/u/oaisoftwarealliance}{{OSA} docker
  repositories} (Nov. 2024).
\newline\urlprefix\url{https://hub.docker.com/u/oaisoftwarealliance}

\bibitem{gemmi2024open6goticblueprintprogrammable}
G.~Gemmi, M.~Polese, P.~Johari, S.~Maxenti, M.~Seltser, T.~Melodia,
  \href{https://arxiv.org/abs/2409.02237}{{Open6G OTIC: A Blueprint for
  Programmable O-RAN and 3GPP Testing Infrastructure}} (October 2024).
\newline\urlprefix\url{https://arxiv.org/abs/2409.02237}

\bibitem{villa2024x5gtmc}
D.~Villa, I.~Khan, F.~Kaltenberger, N.~Hedberg, R.~S. da~Silva, S.~Maxenti,
  L.~Bonati, A.~Kelkar, C.~Dick, E.~Baena, J.~M. Jornet, T.~Melodia, M.~Polese,
  D.~Koutsonikolas, {X5G: An Open, Programmable, Multi-vendor, End-to-end,
  Private 5G O-RAN Testbed with NVIDIA ARC and OpenAirInterface},
  arXiv:2406.15935 [cs.NI] (2024) 1--15.

\bibitem{OAI_jenkins}
{OpenAirInterface Software Alliance},
  \href{https://jenkins-oai.eurecom.fr/}{{OpenAirInterface {CI} pipelines}}
  (2024).
\newline\urlprefix\url{https://jenkins-oai.eurecom.fr/}

\bibitem{arc-ota}
NVIDIA,
  \href{https://docs.nvidia.com/aerial/aerial-ran-colab-ota/current/index.html}{{Aerial
  {RAN} co-lab over-the-air ({ARC-OTA})}} (2024).
\newline\urlprefix\url{https://docs.nvidia.com/aerial/aerial-ran-colab-ota/current/index.html}

\bibitem{Fdida2022}
S.~Fdida, N.~Makris, T.~Korakis, R.~Bruno, A.~Passarella, P.~Andreou,
  B.~Belter, C.~Crettaz, W.~Dabbous, Y.~Demchenko, R.~Knopp,
  \href{https://www.sciencedirect.com/science/article/pii/S0140366422002663}{{SLICES},
  a scientific instrument for the networking community}, Computer
  Communications 193 (2022) 189--203.
\newblock \href {https://doi.org/https://doi.org/10.1016/j.comcom.2022.07.019}
  {\path{doi:https://doi.org/10.1016/j.comcom.2022.07.019}}.
\newline\urlprefix\url{https://www.sciencedirect.com/science/article/pii/S0140366422002663}

\bibitem{slices_5g_blueprint}
{{SLICES} Starting Community},
  \href{https://doc.slices-sc.eu/blueprint/5g_blueprint.html}{{SLICES} {5G}
  blueprint} (2024).
\newline\urlprefix\url{https://doc.slices-sc.eu/blueprint/5g_blueprint.html}

\bibitem{drone4wireless}
{EURECOM}, \href{https://www.drone4wireless.com/home}{{Drone4Wireless} lab}
  (Nov. 2024).
\newline\urlprefix\url{https://www.drone4wireless.com/home}

\bibitem{Gangula2018}
R.~Gangula, O.~Esrafilian, D.~Gesbert, C.~Roux, F.~Kaltenberger, R.~Knopp,
  {Flying Rebots: First Results on an Autonomous {UAV}-Based {LTE} Relay Using
  Open Airinterface}, in: 2018 IEEE 19th International Workshop on Signal
  Processing Advances in Wireless Communications (SPAWC), 2018, pp. 1--5.
\newblock \href {https://doi.org/10.1109/SPAWC.2018.8445947}
  {\path{doi:10.1109/SPAWC.2018.8445947}}.

\bibitem{Mundlamuri2023}
R.~Mundlamuri, O.~Esrafilian, R.~Gangula, R.~Kharade, C.~Roux, F.~Kaltenberger,
  R.~Knopp, D.~Gesbert, \href{https://arxiv.org/abs/2305.05983}{{Integrated
  Access and Backhaul in {5G} with Aerial Distributed Unit using
  OpenAirInterface}} (2023).
\newblock \href {http://arxiv.org/abs/2305.05983} {\path{arXiv:2305.05983}}.
\newline\urlprefix\url{https://arxiv.org/abs/2305.05983}

\bibitem{moro2023nfv}
E.~Moro, M.~Polese, A.~Capone, T.~Melodia, {An Open {RAN} Framework for the
  Dynamic Control of {5G} Service Level Agreements}, in: IEEE Conference on
  Network Function Virtualization and Software Defined Networks (NFV-SDN),
  Dresden, Germany, 2023.

\bibitem{pawr}
{\acrlong{pawr}}, \href{https://www.advancedwireless.org}{{\acrlong{pawr}}
  website}, accessed August 2024 (2024).
\newline\urlprefix\url{https://www.advancedwireless.org}

\bibitem{bonati2021colosseum}
L.~Bonati, P.~Johari, M.~Polese, S.~D'Oro, S.~Mohanti, M.~Tehrani-Moayyed,
  D.~Villa, S.~Shrivastava, C.~Tassie, K.~Yoder, A.~Bagga, P.~Patel, V.~Petkov,
  M.~Seltser, F.~Restuccia, A.~Gosain, K.~R. Chowdhury, S.~Basagni, T.~Melodia,
  {Colosseum: Large-Scale Wireless Experimentation Through Hardware-in-the-Loop
  Network Emulation}, in: Proceedings of IEEE DySPAN, 2021.

\bibitem{villa2023dt}
D.~Villa, M.~Tehrani-Moayyed, C.~P. Robinson, L.~Bonati, P.~Johari, M.~Polese,
  S.~Basagni, T.~Melodia, {Colosseum as a Digital Twin: Bridging Real-World
  Experimentation and Wireless Network Emulation}, arXiv:2303.17063 [cs.NI]
  (2023) 1--16.

\bibitem{polese2024colosseum}
M.~Polese, L.~Bonati, S.~D’Oro, P.~Johari, D.~Villa, S.~Velumani, R.~Gangula,
  M.~Tsampazi, C.~P. Robinson, G.~Gemmi, et~al., {Colosseum: The Open {RAN}
  Digital Twin}, IEEE Open Journal of the Communications Society (2024).

\bibitem{bonati20235gct}
L.~Bonati, M.~Polese, S.~D'Oro, P.~Brach~del Prever, T.~Melodia, {5G-CT:
  Automated Deployment and Over-the-Air Testing of End-to-End Open Radio Access
  Networks}, arXiv:2311.03206 [cs.NI] (2023) 1--7.

\bibitem{POWDER2024}
{Platform for Open Wireless Data-driven Experimental Research},
  \href{https://www.powderwireless.net/5g}{{5G} on powder}, online (2024).
\newline\urlprefix\url{https://www.powderwireless.net/5g}

\bibitem{boateng2024arasdr}
J.~O. Boateng, T.~Zhang, G.~Zu, T.~U. Islam, S.~Babu, H.~Zhang, D.~Qiao,
  {AraSDR}: End-to-end, fully-programmable living lab for {5G} and beyond, in:
  ICC 2024-IEEE International Conference on Communications, IEEE, 2024, pp.
  1758--1763.

\bibitem{zhang2024ara}
T.~Zhang, J.~O. Boateng, T.~U. Islam, A.~Ahmad, H.~Zhang, D.~Qiao, {Ara-O-RAN}:
  End-to-end programmable {O-RAN} living lab for agriculture and rural
  communities, in: IEEE INFOCOM 2024-IEEE Conference on Computer Communications
  Workshops (INFOCOM WKSHPS), IEEE, 2024, pp. 1--6.

\bibitem{OforiBoateng2025}
J.~{Ofori Boateng}, T.~Zhang, G.~Zu, T.~U. Islam, S.~Babu, F.~Kaltenberger,
  R.~Schmidt, H.~Zhang, D.~Qiao, {AraRACH:} enhancing {NextG} random access
  reliability in programmable wireless living labs, in: 2025 IEEE 11th
  International Conference on Network Softwarization (NetSoft) (NetSoft 2025),
  Budapest, Hungary, 2025, p. 8.99.

\bibitem{Moore2024}
J.~Moore, A.~S. Abdalla, C.~Ueltschey, V.~Marojevic,
  \href{https://arxiv.org/abs/2411.04027}{{Prototyping {O-RAN} Enabled {UAV}
  Experimentation for the {AERPAW} Testbed}} (2024).
\newblock \href {http://arxiv.org/abs/2411.04027} {\path{arXiv:2411.04027}}.
\newline\urlprefix\url{https://arxiv.org/abs/2411.04027}

\bibitem{Allbesmart2025FR3}
P.~Marques, L.~Pereira, T.~Alves, A.~Dhananjav, N.~Pandeya,
  \href{https://openairinterface.org/openairinterface-for-fr3-experimentation/}{{OpenAirInterface
  for FR3 experimentation}}, Webinar (Mar. 2025).
\newline\urlprefix\url{https://openairinterface.org/openairinterface-for-fr3-experimentation/}

\bibitem{Gangula2024}
R.~Gangula, A.~Lacava, M.~Polese, S.~D'Oro, L.~Bonati, F.~Kaltenberger,
  P.~Johari, T.~Melodia, {Listen-While-Talking:} toward {dApp-Based}
  {Real-Time} spectrum sharing in {O-RAN}, in: 2024 IEEE Military
  Communications Conference (MILCOM 2024), Washington, USA, 2024, p.~2.

\bibitem{mundlamuri20245g}
R.~Mundlamuri, S.~Badran, R.~Gangula, F.~Kaltenberger, J.~M. Jornet,
  T.~Melodia, {5G} over terahertz using openairinterface, in: 2024 19th
  Wireless On-Demand Network Systems and Services Conference (WONS), IEEE,
  2024, pp. 29--32.

\bibitem{Cheng2024}
H.~Cheng, P.~Johari, M.~A. Arfaoui, F.~Periard, P.~Pietraski, G.~Zhang,
  T.~Melodia, {Real-Time AI-Enabled CSI Feedback Experimentation with Open
  RAN}, in: Proc. of IEEE/IFIP Wireless On-demand Network systems and Services
  Conference (WONS), Chamonix, France, 2024.

\bibitem{Ramantas2023}
K.~Ramantas, A.~N. Bikos, W.~Nitzold, S.~Pollin, A.~Ksentini, S.~Mayrargue,
  V.~Theodorou, L.~Christofi, G.~Gardikis, M.~A. Rahman, A.~Chawla, F.~Ibañez,
  I.~Chochliouros, D.~Nicholson, Mario, Montagudand, A.~Shojaeifard,
  A.~Pagkotzidis, C.~Verikoukis, {6G-BRICKS}: Building reusable testbed
  infrastructures for cloud-to-device breakthrough technologies, in: 2023 IEEE
  Globecom Workshops (GC Wkshps), 2023, pp. 751--756.
\newblock \href {https://doi.org/10.1109/GCWkshps58843.2023.10465223}
  {\path{doi:10.1109/GCWkshps58843.2023.10465223}}.

\bibitem{Bonati2023}
L.~Bonati, M.~Polese, S.~D’Oro, S.~Basagni, T.~Melodia, {OpenRAN Gym}:
  {AI/ML} development, data collection, and testing for {O-RAN} on {PAWR}
  platforms, Computer Networks 220 (2023) 109502.

\bibitem{Boutiba2023}
K.~Boutiba, M.~Bagaa, A.~Ksentini, {On enabling {5G} dynamic {TDD} by
  leveraging deep reinforcement learning and {O-RAN}}, in: NOMS 2023-2023
  IEEE/IFIP Network Operations and Management Symposium, IEEE, 2023, pp. 1--3.

\bibitem{Lacava2025}
A.~Lacava, L.~Bonati, N.~Mohamadi, R.~Gangula, F.~Kaltenberger, P.~Johari,
  S.~D'Oro, F.~Cuomo, M.~Polese, T.~Melodia,
  \href{https://arxiv.org/abs/2501.16502}{{dApps: Enabling Real-Time AI-Based
  Open RAN Control}} (2025).
\newblock \href {http://arxiv.org/abs/2501.16502} {\path{arXiv:2501.16502}}.
\newline\urlprefix\url{https://arxiv.org/abs/2501.16502}

\bibitem{Mohammadi2024}
A.~Mohammadi, N.~Nikaein, {Athena: An Intelligent Multi-x Cloud Native Network
  Operator}, IEEE Journal on Selected Areas in Communications 42~(2) (2024)
  460--472.
\newblock \href {https://doi.org/10.1109/JSAC.2023.3336172}
  {\path{doi:10.1109/JSAC.2023.3336172}}.

\bibitem{Shankaranarayanan2024}
N.~Shankaranarayanan, Z.~Li, I.~Seskar, P.~Maddala, S.~Puthenpura, A.~Stancu,
  A.~Agarwal, {POET: A Platform for O-RAN Energy Efficiency Testing}, in: 2024
  IEEE 100th Vehicular Technology Conference (VTC2024-Fall), IEEE, 2024, pp.
  1--5.

\bibitem{Martins2024}
S.~Martins, A.~Aguiar, P.~Steenkiste, {A Methodology and Testbed to Develop an
  Energy Model for 5G Virtualized RANs}, in: 2024 IEEE International Conference
  on Communications Workshops (ICC Workshops), IEEE, 2024, pp. 982--987.

\bibitem{Centofanti2024}
C.~Centofanti, G.~Venkateswarlu, J.~Santos, R.~R. Tella, A.~Marotta,
  F.~Graziosi, K.~Kondepu, {An Energy Measurement Framework for 5G RAN Using
  USRP and Real-Time Monitoring}, in: 2024 IEEE International Conference on
  Advanced Networks and Telecommunications Systems (ANTS), 2024, pp. 1--5.
\newblock \href {https://doi.org/10.1109/ANTS63515.2024.10898669}
  {\path{doi:10.1109/ANTS63515.2024.10898669}}.

\bibitem{ITU2024}
{International Telecommunication Union},
  \href{https://www.itu.int/dms_pub/itu-r/opb/act/R-ACT-WRC.16-2024-PDF-E.pdf}{{Final
  Acts of the World Radiocommunication Conference (WRC-23)}}, accessed:
  2024-09-30 (2024).
\newline\urlprefix\url{https://www.itu.int/dms_pub/itu-r/opb/act/R-ACT-WRC.16-2024-PDF-E.pdf}

\bibitem{Bjoernson2024}
E.~Björnson, F.~Kara, N.~Kolomvakis, A.~Kosasih, P.~Ramezani, M.~B. Salman,
  \href{https://arxiv.org/abs/2407.05630}{{Enabling 6G Performance in the Upper
  Mid-Band by Transitioning From Massive to Gigantic MIMO}} (2024).
\newblock \href {http://arxiv.org/abs/2407.05630} {\path{arXiv:2407.05630}}.
\newline\urlprefix\url{https://arxiv.org/abs/2407.05630}

\bibitem{3GPP38.843}
3GPP, {Study on Artificial Intelligence ({AI})/Machine Learning ({ML}) for {NR}
  air interface}, techreport 38.843 (Jan. 2024).

\bibitem{Guo2024}
J.~Guo, C.-K. Wen, S.~Jin, X.~Li, {AI} for {CSI} feedback enhancement in
  {5G}-advanced, IEEE Wireless Communications 31~(3) (2024) 169--176.
\newblock \href {https://doi.org/10.1109/MWC.010.2200304}
  {\path{doi:10.1109/MWC.010.2200304}}.

\bibitem{Groen2024}
J.~Groen, S.~D'Oro, U.~Demir, L.~Bonati, D.~Villa, M.~Polese, T.~Melodia,
  K.~Chowdhury, {Securing O-RAN Open Interfaces}, IEEE Transactions on Mobile
  Computing 23~(12) (2024) 11265--11277.
\newblock \href {https://doi.org/10.1109/TMC.2024.3393430}
  {\path{doi:10.1109/TMC.2024.3393430}}.

\bibitem{Wen2024}
H.~Wen, P.~Sharma, V.~Yegneswaran, P.~Porras, A.~Gehani, Z.~Lin, {6G-XSec:
  Explainable Edge Security for Emerging OpenRAN Architectures}, in:
  Proceedings of the 23rd ACM Workshop on Hot Topics in Networks, 2024, pp.
  77--85.

\bibitem{lin2023embracing}
X.~Lin, L.~Kundu, C.~Dick, S.~Velayutham, {Embracing {AI} in {5G}-Advanced
  toward {6G}: A joint {3GPP} and {O-RAN} perspective}, IEEE Communications
  Standards Magazine 7~(4) (2023) 76--83.

\bibitem{osateam}
{OpenAirInterface Software Alliance},
  \href{https://openairinterface.org/oai-team}{{OAI} team} (2024).
\newline\urlprefix\url{https://openairinterface.org/oai-team}

\bibitem{osamembers}
{OpenAirInterface Software Alliance},
  \href{https://openairinterface.org/osa-members/}{{List of Members}} (2024).
\newline\urlprefix\url{https://openairinterface.org/osa-members/}

\end{thebibliography}

\end{document}